\documentclass[twocolumn,superscriptaddress,prd,showkeys,showpacs,nofootinbib]{revtex4-1}

\usepackage{graphicx}
\usepackage[colorlinks = true,
            linkcolor = cyan,
            urlcolor  = blue,
            citecolor = red,
            anchorcolor = blue]{hyperref}\usepackage{color}
\usepackage{amssymb}
\usepackage[nointegrals]{wasysym}
\usepackage{amsthm}
\usepackage{textcomp}
\usepackage{mathtools}

\usepackage{lineno}

\usepackage{comment}
\usepackage[separate-uncertainty,retain-explicit-plus,per-mode = symbol]{siunitx}
\usepackage{url}

\begin{document}

\title{Electron-ion recombination in composite interactions in liquid xenon}

\author{J.~Xu} \email[Corresponding author, ] {xu12@llnl.gov}\affiliation{Lawrence Livermore National Laboratory, Livermore, CA 94551, USA}  
\author{J. Kim} \affiliation{University of California, Santa Barbara, Department of Physics, Santa Barbara, CA 93106-9530, USA}  
\author{B.~Lenardo} \affiliation{SLAC National Accelerator Laboratory, Menlo Park, CA 94025, USA}  
\author{C.E.~Dahl}
\affiliation{Northwestern University, Department of Physics \& Astronomy, Evanston, IL 60208-3112, USA}
\author{R.L.~Mannino} \affiliation{Lawrence Livermore National Laboratory, Livermore, CA 94551, USA}  
\author{G.M.~Blockinger}
\affiliation{University at Albany (SUNY), Department of Physics, Albany, NY 12222-0100, USA}
\author{C.A. Hardy} \affiliation{Physics Department, Stanford University, Stanford, CA 94305, USA}
\author{D.~Adams} \affiliation{C.N. Yang Institute for Theoretical Physics, Stony Brook University, Stony Brook, NY 11794, USA}  
\author{C.S.~Amarasinghe}
\affiliation{University of California, Santa Barbara, Department of Physics, Santa Barbara, CA 93106-9530, USA}
\author{J.~Bang}
\affiliation{Brown University, Department of Physics, Providence, RI 02912-9037, USA}
\author{A.C.~Vaitkus}
\affiliation{Brown University, Department of Physics, Providence, RI 02912-9037, USA}
\author{C.~Ding}
\affiliation{Brown University, Department of Physics, Providence, RI 02912-9037, USA}
\author{W.H.~Lippincott}
\affiliation{University of California, Santa Barbara, Department of Physics, Santa Barbara, CA 93106-9530, USA}
\author{M.~Szydagis}
\affiliation{University at Albany (SUNY), Department of Physics, Albany, NY 12222-0100, USA}
\author{C.~Levy}
\affiliation{University at Albany (SUNY), Department of Physics, Albany, NY 12222-0100, USA}
\author{R.J.~Gaitskell}
\affiliation{Brown University, Department of Physics, Providence, RI 02912-9037, USA}
\author{R.~Essig} \affiliation{C.N. Yang Institute for Theoretical Physics, Stony Brook University, Stony Brook, NY 11794, USA} 

\date{\today}

\begin{abstract}
The response of liquid xenon to various types of ionizing radiation has been extensively studied theoretically and experimentally. Recent progress in direct detection dark matter experiments highlights the significance of composite events, where multiple particles interact with xenon simultaneously and generate overlapping ionization signatures. In these events, recombination of electrons and ions associated with different primary particles leads to additional suppression of the ionization signal, introducing a new source of uncertainty in dark matter searches and Migdal effect studies.
We developed a model to estimate the recombination enhancement for overlapping low-energy particle interactions. This method, which has minimal dependence on xenon microphysics and is primarily driven by existing experimental data, yields predictions that are consistent with available measurements of composite interactions. Furthermore, we demonstrate that the model predictions are robust against xenon microphysics assumptions.
\end{abstract}

\maketitle




\section{Introduction}
\label{sec:intro}

Liquid xenon is widely used in rare event searches because of its low intrinsic radioactivity, the production of both scintillation and ionization signals, and its ability to identify the types of particle interactions from the energy partition between these two signal channels~\cite{Xenon10_ERNR}. 
Currently there are 3 multi-tonne dual-phase xenon Time Projection Chamber (TPC) experiments in operation that search for direct dark matter interactions \cite{LZ2024_WS, XENON2024_8B, PandaX2024_8B}. 
A tonne-scale xenon detector with a $^{136}$Xe-enriched target has also been proposed to study the hypothesized neutrinoless double beta decay process \cite{adhikari2021nexo}. 

The response of liquid xenon to different particle interactions has been extensively studied experimentally, and a comparable effort is made to model these interactions~\cite{DahlThesis,Szydagis2025_NEST}. In particular, the charge and light outputs of liquid xenon from nuclear recoil (NR) interactions, which represent possible dark matter signals, have been calibrated down to the sub-keV region~\cite{Lenardo2019_XeNR,LUX2025_NR} and up to hundreds of keV~\cite{Pershing2022_XeNR}, and its responses to electron recoils (ERs) have also been studied with a variety of sources including $^{37}$Ar, $^3$H, and $^{83m}$Kr~\cite{XENON2023_Ar37, LUX2016_3H, Boulton2017_Ar37, LUX2017_Kr83m}.
These calibrations have enabled the precise modeling of expected dark-matter-induced NR signals in liquid xenon and their strong separation from ambient ER backgrounds. 
The precision of measurement data and model predictions continues to be improved with ongoing efforts.

Recent progress in xenon-based experiments has drawn attention to a new class of particle interactions, which produces multiple ionizing particles at the same time and the same location. 
Due to the close proximity, the electrons and ions produced by different ionizing particles overlap and can interact; as a result, the observable light and charge signals by a particle are altered by neighboring particles. 
We refer to these events as composite interactions. 
Different from a ``multisite'' interaction that contains two or more (nearly) simultaneous energy depositions at well-separated locations, a composite event produces a xenon response that is not the sum of those for the constituent interactions. 

An example of composite interaction is the electron capture (EC) decay, in which a nucleus captures an inner shell electron and emits a neutrino. 
Because neutrinos usually escape a detector without interacting, the main signature of an EC decay is the atomic relaxation of the shell vacancy and, sometimes, deexcitation  gamma rays from the newly formed nucleus. 
EC decays of $^{125}$Xe, $^{127}$Xe and $^{37}$Ar are widely used in the calibration of xenon detectors for their well-defined, low-energy signatures~\cite{LUX2017_Xe127, Temples2021_127Xe, XENON1T_2017,Akimov2014_Ar37,XENON2023_Ar37,Boulton2017_Ar37,LZ2025_EC}. 
However, EC decays usually lead to multiple decay products,  making the xenon response complex. 
For example, the primary decay channel of an L-shell xenon vacancy (or a higher shell) is Auger electron emission. 
Each Auger decay results in the emission of an electron and the production of two outer shell vacancies, which would also decay primarily through the Auger process. 
Therefore, a single xenon L-shell EC may produce several low-energy electrons at nearly the same time. 
The slowdown of each primary particle in liquid xenon produces a group of xenon ions and electrons, 
the recombination of which determines the measurable light and charge signal strengths. 
In an EC decay, the exposure of  xenon ions from one primary particle to electrons from other primary particles leads to additional electron-ion recombination. 
As a result, the signal output of an EC decay can differ from that of a single electron of the same energy. 
This effect was observed in XELDA and LZ, with the L-shell EC decays measured to have $\sim$90\% of the beta charge yield~\cite{Temples2021_127Xe,LZ2025_EC}. 

The charge suppression effect from recombination enhancement is more significant in events with more complex compositions. 
Double EC (DEC) decays of $^{124}$Xe have been observed in the 
large xenon TPC experiments~\cite{Xenon2019_124Xe, LZ2024_124Xe, XnT2022_NewPhysics, Panda2024_124Xe}.
$^{124}$Xe is a naturally occurring xenon isotope with $\sim$0.1\% abundance. It can decay to $^{124}$Te by capturing two inner shell electrons and emitting two neutrinos, with an extraordinarily long half-life of $>$10$^{22}$ years. 
Due to the two simultaneous EC decays at the same location, the light and charge generation process in $^{124}$Xe DEC events is strongly modified. 
In the 2024 WIMP search of LZ (WS2024), the observed charge yield of double-L EC decays is only $\sim$64\% of the sum of two single-L EC decays, or $\sim$70\% of that of beta events of the same total energy~\cite{LZ2024_WS}. 
The reduction in charge yield and increase of light yield cause these events to leak into the NR parameter space more than typical ERs, making them a dangerous background in dark matter searches. 
Similarly, elastic neutrino scattering with electrons in xenon can produce a recoiling electron and a simultaneous atomic vacancy,  another composite event topology with an increased leakage into the dark matter parameter space. 
As liquid xenon experiments continue to increase exposure and grow in size~\cite{XLZD2024_DesignBook}, 
 the rare $^{124}$Xe DEC decays and neutrino-electron scatters will become major backgrounds, and an independent evaluation of the xenon response is required. 

Another important signal with a composite signature is the Migdal interaction. The Migdal effect predicts that when a neutral particle scatters off a nucleus, it could result in an atomic excitation or ionization due to the electromagnetic coupling between the nucleus and shell electrons~\cite{Migdal_1941,Ibe2017_Migdal}. 
This signal contains both an NR and an ER component, and the latter further consists of a Migdal electron (in the case of Migdal ionization) and an atomic shell vacancy. 
The inelastic nature of this process allows a bigger fraction of the neutral particle's kinetic energy to be transferred, which can make a detector more sensitive to low-energy particles through scattering.
Despite the extremely small probability for Migdal interactions~\cite{Ibe2017_Migdal}, 
several experiments have claimed competitive low-mass dark matter sensitivities by assuming this signal channel~\cite{LUX2019:Migdal,CDMS2022_Migdal,XENON1T2019:Migdal,DarkSide2022:Migdal}.
Unfortunately, the effect has not been experimentally verified to date. 
In addition to other challenges, the composite nature of the Migdal signal creates large uncertainties in the signal response~\cite{Xu2024_Migdal,LZ_Bang_phd,LZ_Vaitkus_phd}, 
making a definitive search difficult. 
Even if the Migdal effect is confirmed, dark matter searches with considerable NR energy deposits are still subject to the aforementioned uncertainty in the signal model~\cite{DarkSide2022:Migdal}. 

Direct measurements of liquid xenon's response to composite interactions are highly desired. However, the only measurements demonstrated so far are the simple single-shell EC decays by $^{125}$Xe, $^{127}$Xe, and $^{37}$Ar~\cite{LZ2023_Background, Temples2021_127Xe, XENON2023_Ar37,Boulton2017_Ar37,Akimov2014_Ar37,LZ2025_EC}. 
Calibration of other composite tracks is less practical due to the following reasons. 
For $^{124}$Xe DEC decays, its low isotopic abundance and extraordinarily long half-life make a high-precision measurement prohibitive; so far, the highest statistics for this process come from LZ, an experiment that suffers from a degeneracy between this background and high-energy  signals in some dark matter models~\cite{Fitzpatrick:2012ix}. 
Similarly, despite its rarity, a Migdal interaction is by far the most likely mechanism to produce overlapping NRs and ERs. 
On the other hand, proxy calibrations face substantial difficulty in producing prescribed ionizing particles that perfectly overlap. 
Compton scatter and photoabsorption of gamma-rays can liberate an electron and simultaneously produce an atomic vacancy, but the probability for innershell ionization is low and the signals cannot be conveniently separated from backgrounds. 
It should be noted that regular ``single-track'' particle interactions may contain composite tracks when an energetic secondary particle is produced, but this process is stochastic and cannot be controlled.
In fact, it is one of the main sources of recombination fluctuation~\cite{DahlThesis}, which is captured statistically in the measured spreads of charge and light signals. 

This work aims to provide a model that evaluates the liquid xenon response to low-energy composite events based on its measured response to the constituent interactions. We take a data-driven approach that has a minimal dependence on xenon microphysics. 
The model predictions will be compared with available experimental data from composite interactions. 
This paper is organized as follows: 
Section~\ref{sec:method} describes the method used for the modeling; Sec.~\ref{sec:result} presents the results of the model and their comparison with data; Sec.~\ref{sec:disc} discusses the validity of the assumptions and implementations of the modeling method, and Sec.~\ref{sec:concl} concludes this work. 

\section{Modeling Method}
\label{sec:method}

A composite event contains multiple ionizing particles emitted from the same location at the same time. Each primary particle produces a positive ion track and a negative electron cloud as it slows down in liquid xenon. 
In this work, we use the phrase ``ion track'', or ``track'' in short, to refer to a group of ions affiliated with a primary ionizing particle. 
The movement of positive ions under electric fields or from diffusion is relatively small, 
so they are treated as stationary. 
The electrons that carry significant initial kinetic energy will have the energy dissipated through scattering with neighboring xenon atoms and thermalize~\footnote{In an external electric field electrons are accelerated by the field while they lose energy through collision with atoms, so the ``thermal'' energy distribution depends on the electric field strength.}. 
If an electron wanders close to a positive ion and its kinetic energy does not allow it to escape, it may recombine with the ion. 
A good understanding of the recombination process for single particle tracks in liquid xenon has been achieved by combining experimental measurements and theoretical modeling~\cite{Thomas1987_Recombination,DahlThesis,Szydagis2025_NEST}. 

We extend the knowledge of how an ion track recombines with its ``native'' electron cloud to its interactions with ``foreign'' electrons from a colocated ion track. 
This approach is motivated by the estimation that ionization electrons can travel 4--5 $\mu$m in liquid xenon before they are thermalized~\cite{Mozumder1995_Recombination}, a distance larger than the typical track size of ERs of up to 50 keV or that of NRs up to 100s of keV~\cite{DahlThesis}. Therefore, the distribution of thermalized electrons with different origins should appear nearly identical to the ion tracks. 
Electron-ion recombination occurring prior to thermalization is indistinguishable from direct excitations~\cite{DahlThesis}; the effect of varying exciton-to-ion ratios is a source of systematic uncertainty and will be discussed in Sec.~\ref{sec:result}. 

This work extracts the recombination property of an ion track and applies it to the interaction with all available electrons in a composite event. 
This treatment assumes that the spatial separation between the tracks, as well as the individual track length, is significantly shorter than the electron cloud size, beyond which a more sophisticated model will be needed. 
The probability for an ion track to recombine with electrons has a complex dependence on the track geometry, the ionization density, its relative location within the electron cloud and also external electric fields. 
The instant recombination probability for any individual electron and ion is hard to estimate, but the average recombination probability of all electrons and ions in the track can be empirically calculated from the observed xenon response. 
For a particle track that, by itself, produces $N_e$ electrons and $N_{ph}$ photons, its initial ion number before recombination ($N_i$) can be calculated using the exciton-to-ion ratio ($N_{ex}/N_i$).
\begin{equation}
    N_i = \frac{N_e+N_{ph}}{1+N_{ex}/N_i} \label{eq:ni}
\end{equation}
Then the average recombination probability is 
\begin{equation}
    \bar{r}=\frac{N_i-N_e}{N_i} \label{eq:rp}
\end{equation}
A simple cross-track recombination model can be constructed by assuming that $\bar{r}$ is a constant property of an ion track, independent of the environment that the track is in, so the same $\bar{r}$ value can be applied to both in-track and cross-track electron-ion recombination in composite events.   
In the case of $m$ colocated ion tracks in a composite event, all \{$\bar{r}_k$\} values, where the track index $k=1...m$, can be independently calculated using Eq.~\ref{eq:rp} if single-track yield data are available. 
However, these $m$ probabilities are not independent of each other because one electron cannot combine with multiple ion tracks.
We therefore interpret $\bar{r}_{k}$ as the conditional probability for an electron to combine with track $k$ when the electron does not combine with any other track. This leads to a renormalization of the recombination probability as follows
\begin{eqnarray}
    P_{k} & = & \frac{\bar{r}_{k}}{A}\prod_{k'\neq k} (1-\bar{r}_{k'}) \nonumber \\
    A & = & \prod_{k} (1-\bar{r}_{k}) + \sum_k \bar{r}_{k} \prod_{k'\neq k} (1-\bar{r}_{k'}) \label{eq:pr}
\end{eqnarray}

Here $P_{k}$ is the normalized probability for an electron to recombine with track $k$, and the intermediate parameter $A$ provides the normalization. 
With all \{$P_{k}$\} values obtained, the multitrack electron-ion recombination process can be modeled as a multinomial distribution, and the number of electrons surviving recombination, or the charge yield of the composite event, can be calculated. 
Because this method models a composite event based on individual tracks under the same condition, the effects of track topologies and external electric fields are automatically accounted for. 

This recombination method is simple to understand and easy to implement, but can overestimate the electron-ion recombination strength due to the oversimplified constant-$r$ assumption. 
The ability of an ion track to capture electrons depends on the number of available ions in the track so the average probability $\bar{r}$ is a function of both the initial track ion number and the final ion number. 
In a composite event, the additional cross-track recombination further reduces the final ion count in a track, 
which also reduces the average recombination probability $\bar{r}$. 
Therefore, results from this constant-$r$ method should be seen as a lower bound of the charge yield for a composite interaction. 

This weakness of the model can be mitigated by incorporating an electron recombination probability that varies with the instant track ion number. 
This approach will be referred to as the running-$r$ method. 
For single-track events, Eq.~\ref{eq:ni} and \ref{eq:rp} calculate the initial ion count, $N_i$, and the average recombination probability, $\bar{r}$, using the observed light ($N_{ph}$) and charge yields ($N_{e}$). 
In the same way, an empirical function of $\bar{r}(N_i)$ can be derived with experimental data from single-track events of different energies.
This approach assumes that the energy dependence of recombination is driven by the ion number in the event, which is approximately true for a modest change of event energy. 
An alternative way to get $\bar{r}(N_i)$ is to use the box recombination model developed by Thomas and Imel (TIB) \cite{Thomas1987_Recombination}. 
TIB predicts an analytical formulation for $\bar{r}$ as:  
\begin{equation}
    \bar{r}=1-\frac{ln(1+\xi)}{\xi}, \; \xi=\frac{N_i \alpha}{4 a^2 \mu_{-} E}\label{eq:tib}
\end{equation}
where $\xi$ is an intermediate parameter, $\alpha$ is the recombination coefficient (recombination probability per ion per unit time), $a$ is a dimensional parameter, $\mu_{-}$ is the electron mobility, and $E$ is the strength of an external electric field.
This model has been used to describe the recombination process in low-energy NR and ER events in liquid xenon~\cite{DahlThesis,Szydagis2025_NEST}. 
Both the empirical and TIB recombination functions are studied in this work. 

\begin{figure}[!t]
    \centering
    \includegraphics[width=0.33\textwidth]{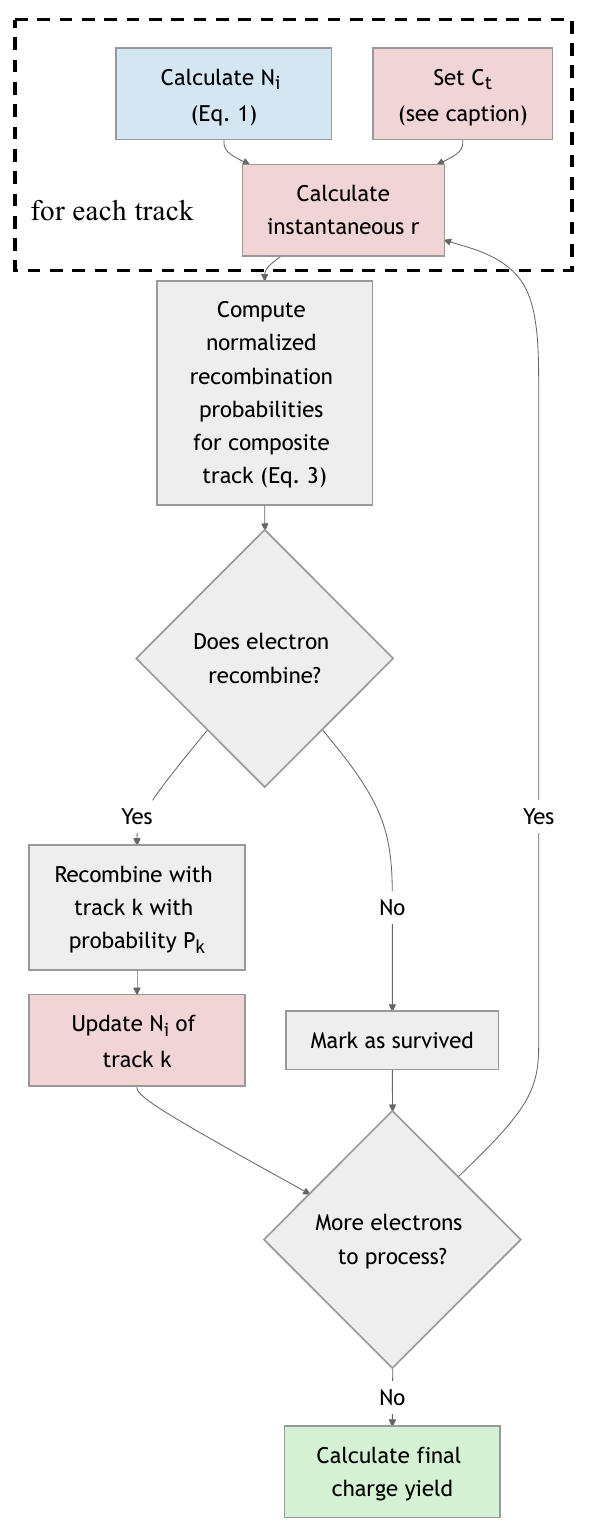}
\caption{Illustration of the running-$r$ recombination modeling process. The dashed box highlights actions on each constituent of a composite track: $N_i$ is calculated from track yields; $\mathcal{C}_t$ is obtained by testing a range of $C_t$ using the 1-track modeling process and finding the value that produces the correct track yield.  The constant-$r$ process differs in the red boxes: instead of $\mathcal{C}_t$, the average $\bar{r}$ (Eq.~\ref{eq:rp}) is used in each subsequent loop until $N_i$ decreases to 0 (then $r=$0).}
    \label{fig:runr}
\end{figure}

In the running-$r$ implementation, we assume that $r(N_i)$, the instant recombination probability for one electron when there are $N_i$ ions in the track, follows the same functional form as $\bar{r}(N_i)$, up to a scaling factor.  
\begin{equation}
  r(N_i) = \bar{r}(\mathcal{C}_t N_i)
\end{equation}
The scaling parameter $\mathcal{C}_t$  is needed because $\bar{r}(N_i)$ is the average recombination probability for all electrons over a range of ion counts in the track, 
which differs from the instant $r(N_i)$ value at the beginning of the process. 
Thanks to the monotonic nature of $\bar{r}(N_i)$, $\mathcal{C}_t$ allows $r(N_i)$ to take a value greater than $\bar{r}(N_i)$, and as the number of ions decreases $r$ is gradually reduced to below the average recombination probability. 
In the TIB case, $\mathcal{C}_t$ also absorbs the track property parameter $\alpha/(4 a^2 \mu_{-} E)$. 
The value of $\mathcal{C}_t$ for a track can be determined from experimental data through Monte Carlo simulations of the recombination process by testing a range of $\mathcal{C}_t$ values. 
Each $\mathcal{C}_t$ corresponds to a unique running-$r$ function and 
produces a distribution of the surviving electron number for the single-track event of interest; 
the $\mathcal{C}_t$ value producing a mean charge yield that matches experimental data 
is chosen and applied to this track in the modeling of composite interactions. 
When $r(N_i)$ is applied to the tracks in a composite event, 
the probability renormalization and recombination simulation follow the same approach as used in the constant-$r$ method. 
This process is illustrated in Fig.~\ref{fig:runr}. 

However, the interpretation of the simulation step has additional subtleties. 
In the constant-$r$ method, because the track property does not change between electrons, the order of electrons in the simulation does not matter, and an electron-by-electron simulation is equivalent to a method that treats recombination of all electrons simultaneously. 
In the running-$r$ method, the electron-by-electron simulation indicates that recombination occurs sequentially. 
Under this approach, the ions that have recombined with electrons in the early stage of the simulation have absolutely no effect on electrons considered later in the simulation. 
In reality, the recombination process for all electrons occur simultaneously.
Every electron is initially subject to the Coulomb force of all ions, and then the number of relevant ions gradually decreases until the electron of interest recombines or escapes. 
By simplifying the simultaneous multibody recombination process to a sequential single-electron process, this method underestimates the recombination probability. 
This effect is compensated for in the running-$r$ simulation of single-track events by tuning $\mathcal{C}_t$ to anchor the model prediction to experimental data from single-track events, but the compensation falls short in composite events in which the ion counts in the tracks are further reduced. 
As a result, the running-$r$ method carries a small underestimate of the recombination probability and the results should be seen as an upper bound of the charge yield for composite events. 

\section{Simulation and Results}
\label{sec:result}

In this section, we apply the developed recombination model to three composite interaction examples and compare model predictions with experimental measurements. 
The model can be made fully data-driven, using only experimental data from single-track events as input, but when necessary or convenient we also use xenon yields evaluated by the public version of NEST~\cite{Szydagis2025_NEST}, which is a fit to large ensembles of experimental data and provides xenon response as a continuous function of event energy and electric field. 
The uncertainty in NEST-predicted ER yield values is estimated to be 10--15\%\cite{Szydagis2025_NEST}. 

We first study the double L-shell EC decay of $^{124}$Xe, which is a relatively simple example of composite events with two identical tracks. 
At the LZ WS2024 electric field of 96.5 V/cm, the photon and charge yields for single beta particles at the average xenon L-shell EC energy of 5.2 keV are evaluated to be 41.8 ph/keV and 31.9 e-/keV by the default NEST version. 
Due to the composite nature of EC decays, the L-shell EC charge yield is suppressed to $\sim$90\% of that of betas, leading to a measured value of 27.8$\pm$1.0 e-/keV in LZ~\cite{LZ2025_EC}. 
With an exciton-to-ion ratio ($N_{ex}/N_i$) of 0.06 and a $W$-value of 13.5 eV assumed for ER events~\cite{Takahashi1975_Xe,Doke2002_ArXe,DahlThesis}, the initial number of ions prior to recombination in a L-shell EC track is estimated to be 363, and the average electron-ion recombination probability is calculated using Eq.~\ref{eq:rp} to be $\bar{r}$=0.60. 

\begin{figure}[!t]
    \centering
    \includegraphics[width=0.49\textwidth]{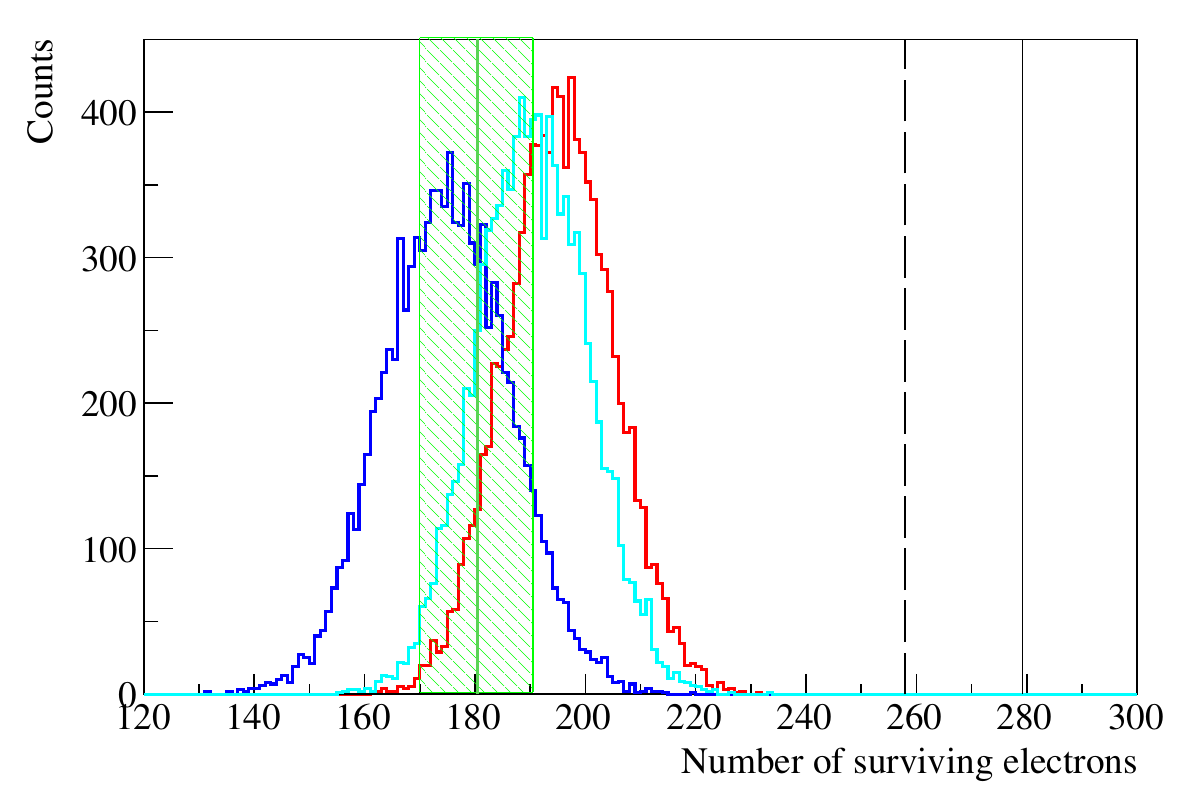}
    \caption{Charge yield distributions predicted for double-L EC decays of $^{124}$Xe estimated using the constant-$r$ (blue), TIB-based running-$r$ (red), and NEST-based running-$r$ (cyan) methods; a $N_{ex}/N_i$ value of 0.06 is assumed. The 1$\sigma$ band of LZ best fit is shown in the green shaded region; the charge yield of 10.04 keV electrons is shown as the vertical dashed line, and the sum of two L-shell EC decays without considering cross-track combination is shown as the vertical solid line. }
    \label{fig:ll}
\end{figure}

The single-L EC recombination probability $\bar{r}$ is then used to simulate the recombination process for the two L-shell tracks in a $^{124}$Xe DEC decay. 
The simulation begins with two ion tracks and two electron clouds; 
and then each electron is tested for recombination using the renormalized probability (Eq.~\ref{eq:pr}). 
It is noteworthy that DEC decays of xenon produce a different nucleus from single EC decays and thus the atomic energy levels are slightly shifted. 
We follow the calculation in \cite{Nitescu2024_124Xe} that predicts a double-L EC decay energy of 10.04 keV and adjust the initial ion numbers accordingly, but assume that the track recombination probability remains the same. 
10,000 simulations are carried out 
to produce a distribution of the number of electrons surviving both in-track and cross-track recombination, as shown in Fig.~\ref{fig:ll} (blue histogram). 
The predicted mean charge yield of 174$\pm$10 electrons is slightly lower than the LZ best fit of 180$\pm$11 electrons~\cite{LZ2024_WS} but the two values agree within uncertainties. 
The uncertainty on this model prediction only includes that from the measured yield for single-L EC decays, and effects of the assumed $N_{ex}/N_i$ value will be discussed later in this section. 

The modeling process with the running-$r$ method requires a few extra steps. 
For this study we tested both an empirical $\bar{r}(N_i)$ function and the TIB function to model the dependence of the recombination probability on the track ion number. 
In principle, the empirical $\bar{r}(N_i)$ function can be derived exclusively using experimental yield data, but for convenience, we calculate it using NEST-evaluated beta yields at the LZ WS2024 electric field. 
With either the empirical or TIB function, we first determine the model parameter $\mathcal{C}_t$ in $r(N_i)$ by simulating the recombination process in a single-L EC decay with a running probability. 
In the simulation, all available electrons are tested for recombination sequentially; if an electron recombines, the track ion count is reduced by one and the track recombination probability is reevaluated based on the test $\mathcal{C}_t$ value; after all electrons are tested, an estimate of the surviving electron number is obtained. 
For each test $\mathcal{C}_t$, 10,000 simulations are carried out to produce a charge yield distribution. 
Then the $\mathcal{C}_t$ value that makes a prediction best matching the LZ measurement is chosen for the simulation of $^{124}$Xe DEC decays. 

The simulation of the two colocated L-shell tracks in a $^{124}$Xe DEC decay is similar to the above process except that there are two initially identical ion tracks and twice more electrons. 
As a result, every time an electron recombines with an ion, the corresponding track's recombination probability is updated and both track's probabilities need to be renormalized. 
The predicted charge yield distributions for double-L EC decays of $^{124}$Xe using the running-$r$ method are also shown in Fig.~\ref{fig:ll}, with the empirical $\bar{r}(N_i)$ result in cyan and the TIB-based result in red. 
The mean charge yields are found to be 190$\pm$11 and 195$\pm$11 electrons, respectively, also consistent with the LZ result but slightly higher. 

It should be noted that when cross-track recombination is not considered, two L-shell EC decays, with the summed energy scaled to that of double-L EC decays of $^{124}$Xe, would produce $\sim$280$\pm$10 electrons (solid vertical line in Fig.~\ref{fig:ll}). The different implementations of our recombination model, using $\sim$700 electrons as the starting point, produce charge yield predictions around 180 electrons near that observed by LZ. 
The predicted charge yield values are also significantly lower than the NEST-predicted yield of 10.04 keV beta particles (dashed vertical line in Fig.~\ref{fig:ll}), 
even when a conservative 15\% NEST yield uncertainty is considered. 
The convergence of the different models and the excellent agreement between model predictions and the LZ measurement strongly support the validity of our proposed recombination enhancement model. 

\begin{figure}[!t]
    \centering
    \includegraphics[width=0.45\textwidth]{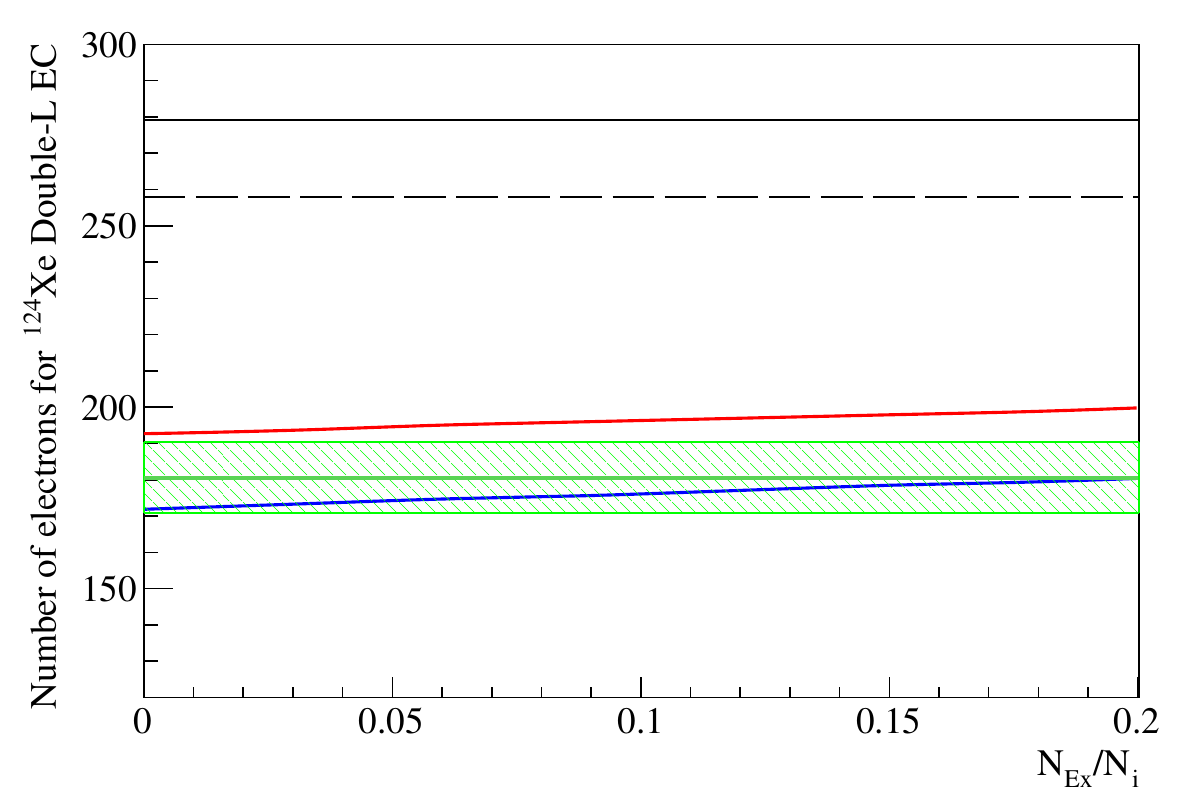}
    \caption{Mean charge yield predicted for double-L EC decays of $^{124}$Xe, with different $N_{ex}/N_i$ assumptions, estimated using the constant-$r$ (blue) and TIB-based running-$r$ (red) methods. The 1$\sigma$ band of LZ best fit is shown in the green shaded region; the charge yield of 10.04 keV electrons (dashed line) and the sum of two L-shell EC decays without considering cross-track combination (solid line) are shown as references.  }
    \label{fig:limits}
\end{figure}

All model implementations have an apparent dependence on $N_{ex}/N_i$ for single-L EC tracks, which has been assumed to be 0.06~\cite{Takahashi1975_Xe,DahlThesis}. However, other estimates of $N_{ex}/N_i$ exist in the literature~\cite{Doke2002_ArXe}, and NEST further models it as energy dependent with a value between 0 and 0.2~\cite{Szydagis2025_NEST}.  
Figure~\ref{fig:limits} shows the $^{124}$Xe DEC yield predictions for $N_{ex}/N_i$ values in this range, with the constant-$r$ result shown in blue and the TIB-based running-$r$ result in red.  
For both methods, varying $N_{ex}/N_i$  only changes the model predictions by $\sim$4\%, which also remain in good agreement with the LZ best fit. 
This is due to our approach of fixing the model predictions for constituent tracks at the experimental observed yields. 
A higher value of $N_{ex}/N_i$ reduces the number of initial ions in a track prior to recombination, and thus requires a weaker recombination strength for the track, which, when applied to double-L EC decays, also partially compensates for the smaller initial ion numbers. 
It is worth noting that reducing the liquid xenon $W$-value has a similar effect to decreasing $N_{ex}/N_i$, and choosing $W$=11.5 eV~\cite{Baudis2021_WValue,EXO2020_WValue} lowers the model predictions by $<$4\%. 
In addition to reducing the model's dependence on the assumed $N_{ex}/N_i$ and $W$ values, 
anchoring single-track predictions to data also mitigates inaccuracies in the chosen $\bar{r}(N_i)$ functions. 
This is reflected in the excellent agreement between the running-$r$ results using the empirical $\bar{r}(N_i)$ function and TIB, as illustrated in Fig.~\ref{fig:ll}. 
For this reason, only the more convenient TIB function is used for the running-$r$ test here and for the rest of this work. 

Next we apply the recombination model to the EC decays of $^{37}$Ar in xenon.
As explained in Sec.~\ref{sec:intro}, each inner shell EC decay is a composite event, 
and the response of liquid xenon to K- and L-shell $^{37}$Ar EC decays has been measured in multiple experiments~\cite{Akimov2014_Ar37,Boulton2017_Ar37,XENON2023_Ar37}. 
However, our model does not support a simulation of the full EC decay process. 
First, each EC decay produces several Auger electrons, and according to EADL~\cite{perkins1991eadl}, most electrons have energies below 100 eV where the yield values are not known. 
Second, the recombination properties of particle tracks containing a small number of ions may depend strongly on the number of ions, which renders them unsuitable for extrapolation to cross-track recombination. 
Third, by simplifying a composite event as a superposition of constituent particle tracks, we treat recombination as a linear process, which may or may not be valid when the number of tracks becomes large. 
Instead, we simulate the K-shell EC decays to L-shell and use L-shell yield measurements to predict the K-shell charge yield.

\begin{figure}[!t]
    \centering
    \includegraphics[width=0.49\textwidth]{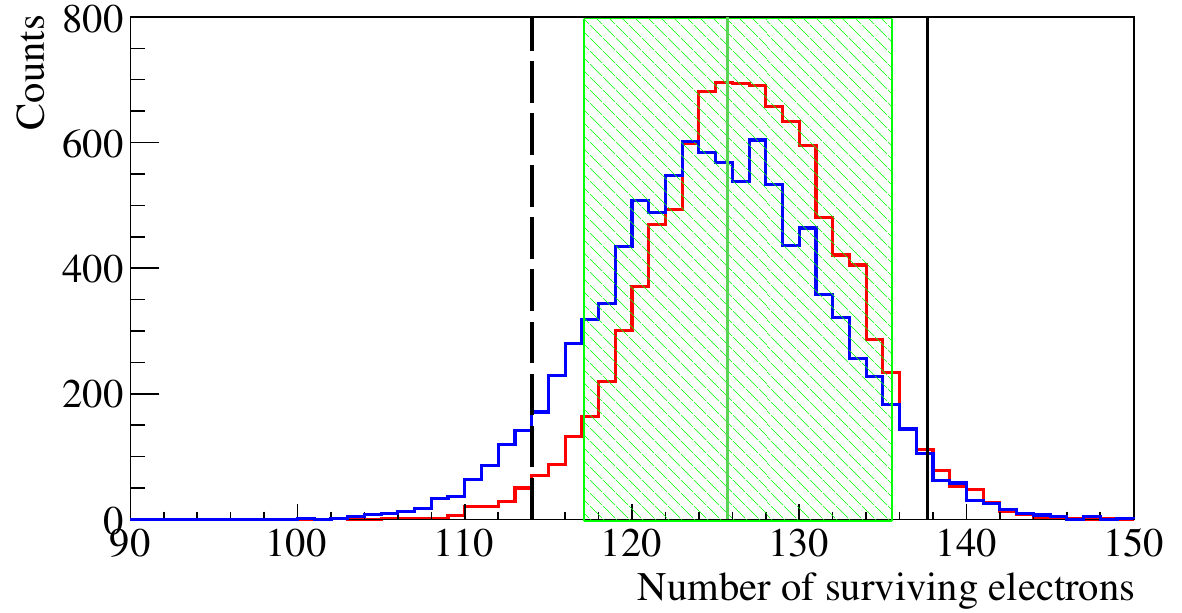}
    \includegraphics[width=0.49\textwidth]{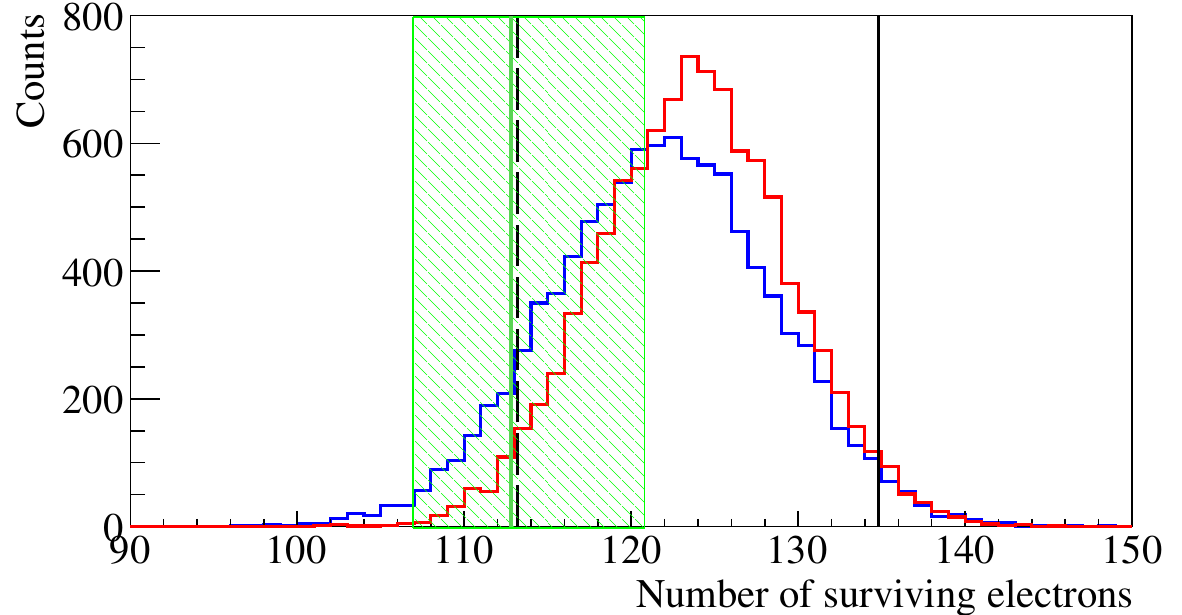}
    \caption{The simulated charge yields of K-shell $^{37}$Ar decays using the constant-$r$ (blue) and TIB-based running-$r$ (red) methods, along with their comparison with PIXeY (top, 100 V/cm) and XENON1T measurements (bottom, 82 V/cm); the models use the measured xenon response to L-shell $^{37}$Ar decays and NEST predicted beta yields as the inputs. The K-shell measurements are shown as the green shaded region; the charge yield of a 2.8 keV electron is shown as the vertical dashed line; the sum of charge signals of constituent tracks without considering cross-track combination is shown as the vertical solid line.}
    \label{fig:ar37}
\end{figure}

Due to the relatively simple atomic structure of $^{37}$Ar, EADL predicts that most K-shell vacancies ($\sim$80\%) decay to two L-shell vacancies (0.2 or 0.27 keV) by emitting an Auger electron of 2.3--2.38 keV energy \cite{perkins1991eadl}. 
As a result, we only simulate the decays of a 2.8 keV K-shell $^{37}$Ar vacancy to a 0.2 keV L-shell vacancy, a 0.27 keV L-shell vacancy, and a 2.33 keV Auger electron. 
In the simulation, we use the L-shell EC yield measured by PIXeY at 100 V/cm~\cite{Boulton2017_Ar37} and that by XENON1T at 82 V/cm~\cite{XENON2023_Ar37}; the Auger electron yield is evaluated by NEST at the corresponding electric field. 
Figure~\ref{fig:ar37} shows the $^{37}$Ar K-shell charge yields predicted with the constant-$r$ method (blue) and the TIB-based running-$r$ method (red) for both PIXeY (top) and XENON1T (bottom).  
The predictions using the two methods are found to be very close to each other, possibly a result of low cross-track recombination due to the small recombination probabilities for single-track ERs at low energies. 
The results are in excellent agreement with PIXeY data but are slightly higher than XENON1T data. 
The discrepancy is mainly attributed to uncertainties in the  K-shell yields measured in the two experiments, which differ by $>$10\% at comparable fields; simplifications in the model such as ignoring the decay modes involving X-ray emission or outer shells may have also contributed additional inaccuracies in the prediction. 
As indicated in Fig.~\ref{fig:ar37}, the model predictions at these two electric fields are similar, but the measured K-shell charge yield values in the two experiments are $>$1$\sigma$ apart. 
A $N_{ex}/N_i$ value of 0.06 is assumed for the ER events in this example, and varying $N_{ex}/N_i$ between 0 and 0.2 for all constituent tracks changes the final results by $\sim$2\%.

Lastly, we study the recombination enhancement of colocated NR and ER tracks as in Migdal signals.
The ER component in a typical Migdal interaction consists of an atomic vacancy and a Migdal ionization electron.
As discussed earlier in this section, the relaxation of an atomic vacancy produces a composite interaction, 
and a suppressed charge yield of $\sim$90\% relative to that of beta particles has been measured for L- and M-shell EC decays of xenon~\cite{Temples2021_127Xe,LZ2025_EC}.
Although EC decays favor inner shell electrons and the Migdal process favors outer shell electrons, the additional Migdal electron makes the average M-shell Migdal ER energy approximately the same as an M-shell EC decay. 
Therefore, we conveniently approximate M-shell Migdal ER signals as monoenergetic EC events at 1.1 keV with yields measured by LZ~\cite{LZ2025_EC}. 
We consider NR energies up to 100 keV, which is slightly above the end point of DD neutron scatters with xenon (74 keV). 
At each NR energy of interest, the in-track and cross-track electron-ion recombination is simulated for 
monoenergetic NRs and 1.1 keV Migdal ERs to estimate the total charge yield. 
We note that the $N_{ex}/N_i$ value for NRs is different from that of ERs, and has been estimated to be 0.7--1.1~\cite{DahlThesis,Szydagis2025_NEST}.
For NRs below 5 keV, the charge and light yields evaluated by NEST may lead to zero or even negative recombination probability for some $N_{ex}/N_i$ values~\footnote{A negative recombination probability is obtained when the initial track ion number estimated using $N_{ex}/N_i$ (Eq.~\ref{eq:ni}) is smaller than the observed charge yield. }, in which cases we use the NEST charge yield as the initial ion number and assume no recombination. 
An electric field of 200V/cm is used so the model prediction can be directly compared to data from the Migdal effect search in liquid xenon at LLNL~\cite{Xu2024_Migdal}. 

\begin{figure}[!t]
    \centering
    \includegraphics[width=0.47\textwidth]{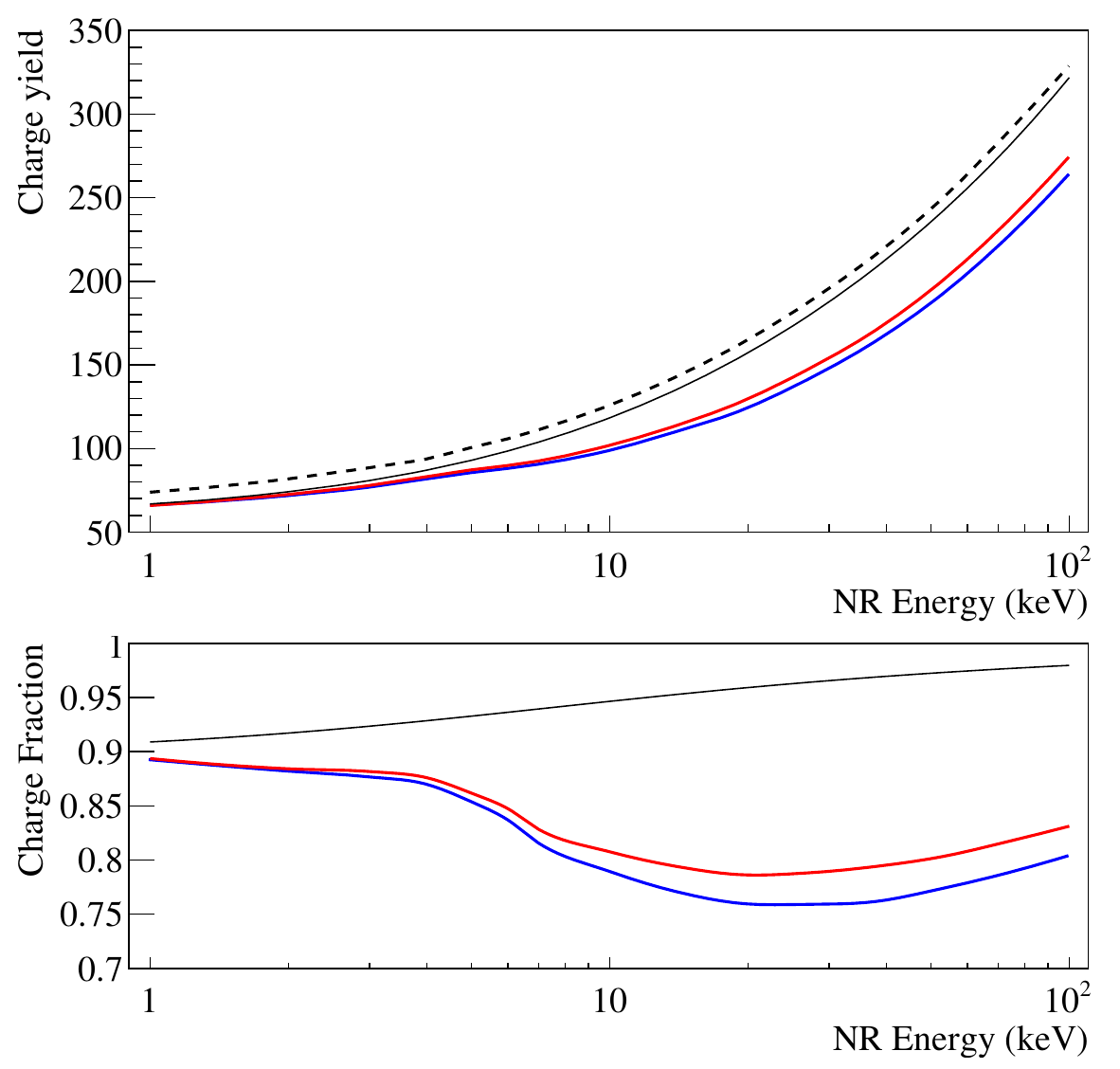}
    \caption{{\bf Top:} the average number of electrons surviving recombination between a 1.1 keV M-shell Migdal ER and monoenergetic NRs, simulated using the constant-$r$ (blue) and running-$r$ (red) methods, in comparison with the simple sum of the signals without considering cross-track recombination, with full (dashed black) and 90\% (solid black) beta yield assumed for the Migdal ER component. {\bf Bottom:} fractions of the evaluated charge yields relative to the dashed black line in the top figure. An electric field of 200 V/cm is used. }
    \label{fig:nr}
\end{figure}

Figure~\ref{fig:nr} (top) shows the predicted charge yields for NR+ER signals using the constant-$r$ (blue) and TIB-based running-$r$ (red) methods. 
The two methods produce similar results with a typical difference of $\sim$2\%. 
The predictions are compared to the sum of the NR and ER responses of liquid xenon (ignoring cross-track recombination) with (solid black) and without (dashed black) considering charge suppression in the ER component. 
Normalized to the NR+ER sum without considering any recombination enhancement, the fractional charge yield slowly decreases with energy to a minimum of $\sim$80\% between 20 keV and 30 keV. 
At high NR energies the charge suppression becomes less significant because the ER component and its effect on the total charge output are relatively small. 
At very low NR energies, the NR component is small relative to the ER which also reduces the effect of cross-track recombination.  
Varying the $N_{ex}/N_i$ value for the NR (ER) component in the full range changes the charge yield estimate by 9\% (4\%) at 7 keV and by $<$1\% (3\%) at 100 keV. 

\begin{figure}[!t]
    \centering
    \includegraphics[width=0.49\textwidth]{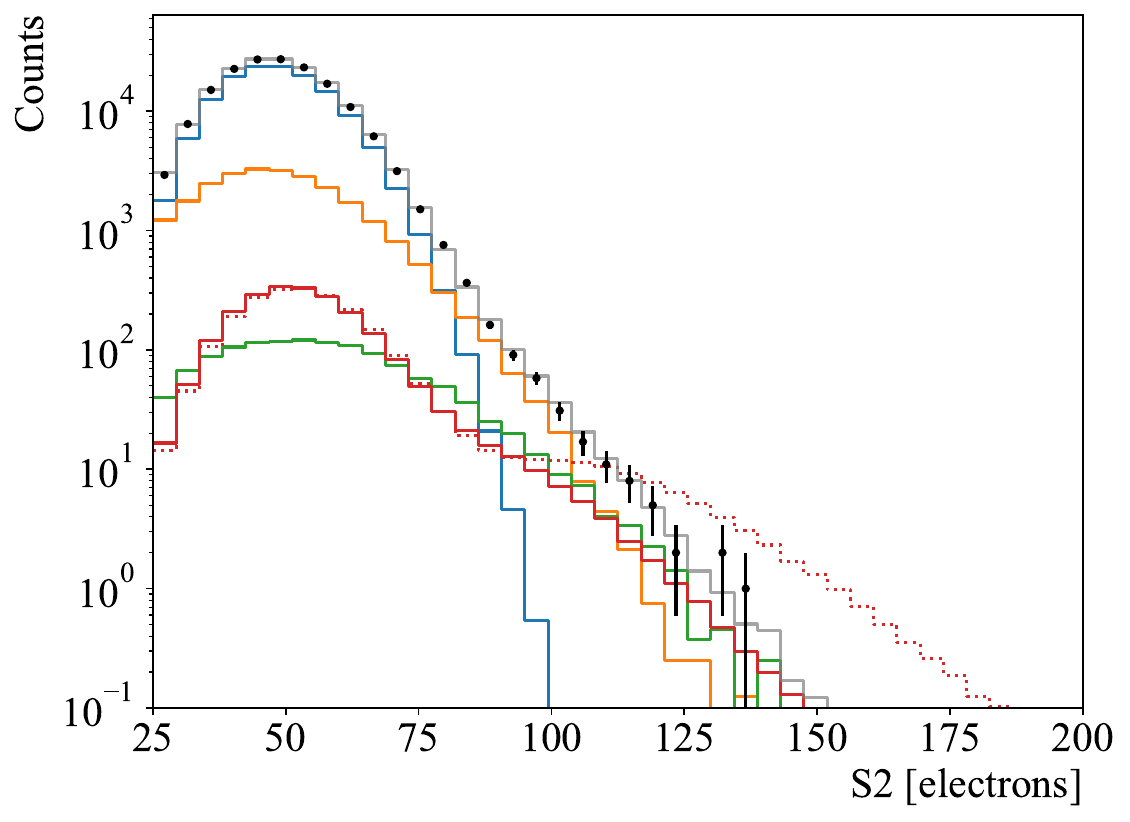}
    \caption{Reanalysis of Migdal search data in \cite{Xu2024_Migdal} using a Migdal signal yield from this work. The electron spectrum with scintillation signals between 4 and 10 photoelectrons is compared with the best-fit signal and background models: Migdal signal at the nominal strength (solid red), single-scatter NR (blue), neutron multiscatters in xenon (green) and in passive materials (orange), and the sum of all components (grey); the Migdal signal model that simply adds the NR and ER yields without considering  recombination enhancement is also shown (dotted red).}
    \label{fig:migdal}
\end{figure}

We then apply the charge yield prediction to the 7.0$\pm$1.5 keV xenon recoil data used to search for the Migdal effect~\cite{Xu2024_Migdal}. 
Following the approach in \cite{Xu2024_Migdal}, the pure NR and ER yields are estimated using NEST, with the NR model parameters tuned to single-scatter NR data in the Migdal search, and the ER parameters tuned to in-situ gamma calibrations. 
The Migdal ER model is separately tuned to produce a lower charge yield so the combined NR+ER yield agrees with model predictions in Fig.~\ref{fig:nr} at 7 keV. 
Figure~\ref{fig:migdal} shows the best fit of the signal and background models to experimental data, with the Migdal signal strength fixed at the theoretical value. 
The suppressed charge yield in the updated Migdal model (solid red), relative to that without considering enhanced recombination (dotted red), causes Migdal events to become degenerate with neutron multiscatter backgrounds in the experiment. 
As a result, the tension between the Migdal prediction and the experimental observation is relieved, confirming that enhanced recombination between colocated NR and ER tracks is a likely explanation of the null Migdal result observed in \cite{Xu2024_Migdal}. 

The recombination model developed in this work will play a critical role in future experimental efforts searching for the Migdal effect in liquid xenon. 
A promising Migdal search approach is to use very low energy NRs. 
As shown in Fig.~\ref{fig:nr}, the NR+ER recombination enhancement is 
$<$5\% below 2 keV, where the modeling uncertainty is also negligible as the 
blue and red curves coincide. 
As a result, the modeling uncertainty in this regime may also be neglected, 
so such Migdal searches will bear the least uncertainty from the cross-track recombination effect. 
In addition, the constant-$r$ method, through its overestimation of recombination enhancement, can guide a Migdal search plan toward a more conservative direction and make a signal claim more convincing. 
This work also provides a more robust framework for modeling signals in dark matter searches using the Migdal effect. 

\section{Discussion}
\label{sec:disc}

The model developed in this work is tested against experimental data containing composite interactions. 
Although results of the constant-$r$ and running-$r$ methods are interpreted as lower and upper bounds on the charge yield of composite events, the predictions are found to converge in these examples and provide strong constraints. 
In this section, we discuss the subtleties in the model that are not covered in Sec.~\ref{sec:method}. 

\begin{figure}[!t]
    \centering
    \includegraphics[width=0.43\textwidth]{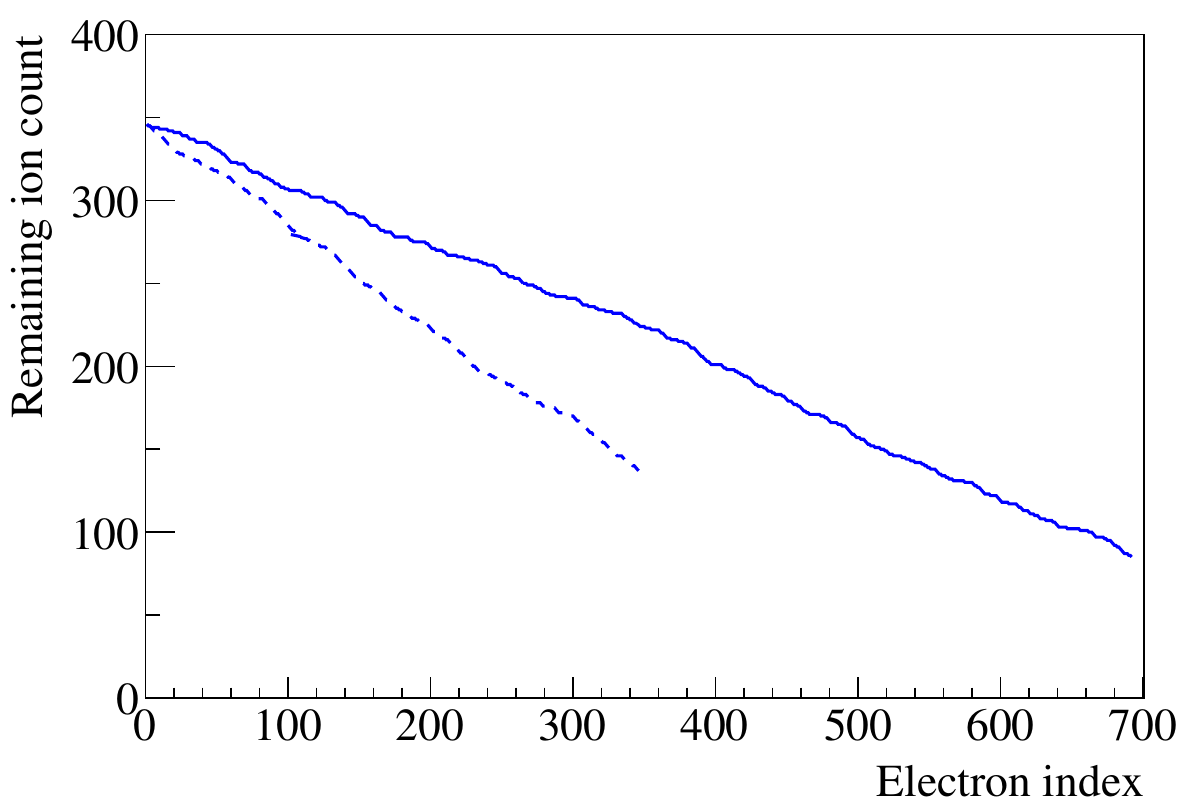}
    \caption{The simulated reduction of ion number in one track of a double-L EC decay of $^{124}$Xe (solid line) and that in a single-L EC decay (dashed line) as electron-ion recombination progresses using the constant-$r$ method. 
    The X axis is the index of electrons tested for recombination; the initial ion and electron numbers in the single-L simulation are scaled to half of the double-L case for direct comparison. }
    \label{fig:ion}
\end{figure}

The constant-$r$ method can overestimate recombination by ignoring the reduction of recombination probability for an ion track in composite interactions. 
Cross-track recombination allows more electrons and ions to recombine, 
and, as the number of ions in the track decreases, its ability to recombine with additional electrons is reduced. 
Therefore, the overall recombination probability for a track should be lower in a composite event than by itself. 
However, this effect is relatively small in the examples studied. 
Figure~\ref{fig:ion} shows the reduction of ion counts in one of the two tracks in the DEC decay simulation of $^{124}$Xe (solid line), in comparison with that in a single-L EC decay with no second track (dashed line). 
Although both simulations start with the same track recombination probability, the ion number in the double-L EC decay does not decrease as fast as that in the single L EC decay. 
This is because each electron is competed for by both ion tracks, which reduces the chance for the electron to recombine with either, as reflected by the probability renormalization in Eq.~\ref{eq:pr}. 
As a result, 
in both scenarios $\sim$100 ions are left in the track after all electrons available are tested for recombination. 
Because the overall track recombination probability is approximately related to the average ion count throughout the recombination process, the effective $\bar{r}$ value in $^{124}$Xe DEC decays would only be modestly reduced from that in single-L EC decays. 

The running-$r$ method incorporates the dependence of the recombination probability on the ion count in a track. 
However, the use of the instant ion count when the electron of interest recombines or escapes neglects the history of the track, and underestimates the recombination probability. 
In single-track events this effect is compensated for by tuning the model parameter $\mathcal{C}_t$ to anchor the model prediction at experimental observation. 
In composite interactions, however, the further reduction of ion count from cross-track recombination is undercompensated. 
This effect causes the recombination probability evaluated by the running-$r$ method to be an underestimate and renders the prediction an upper limit on the charge yield of a composite event. 
Similar to the constant-$r$ method, the bias in the running-$r$ prediction also depends on the additional reduction of ions in composite events (Fig.~\ref{fig:ion}). 
It is worth noting that a $^{124}$Xe DEC decay consists of two identical tracks with high recombination probabilities, so cross-track recombination causes a relatively large change in the final ion count of each track, which leads to a significant difference in the predicted charge yields using the constant-$r$ and running-$r$ methods (Fig.~\ref{fig:ll}). 
When two tracks with a large disparity in ion numbers colocate, cross-track recombination may introduce a larger inaccuracy for the smaller track, but its small contribution to the total signal can mitigate the overall effect. 
In addition, tracks with low ion counts in liquid xenon, such as very low-energy ERs and NRs, usually have low recombination probabilities which helps limit the extra loss of ions to cross-track recombination. 
For this reason, the model predictions in these cases may provide tighter bounds, as exampled by Fig.~\ref{fig:ar37} and Fig.~\ref{fig:nr}. 

In addition to uncertainties from the modeling methods, another significant source of uncertainty is from inaccuracies in the input yields of constituent tracks. 
For example, a 1 e$^{-}$/keV shift in the assumed single-L xenon EC yield will change the predicted charge yield of $^{124}$Xe DEC decays (10.04 keV) by approximately 10 electrons, comparable to the uncertainty on the LZ best fit. 
For the running-$r$ method, an additional uncertainty comes from the function that is used to describe the dependence of the recombination probability on the track ion count. 
Two functions are studied for $^{124}$Xe DEC decays~\footnote{The constant-$r$ method may be seen as a special case where the recombination probability does not vary with the track ion count.}: one derived empirically using the NEST-evaluated beta yields, and the other using the TIB model. 
The close match of the two predictions, as shown in Fig.~\ref{fig:ll}, suggests that the uncertainty on either function is relatively small in this energy region, so the analytical TIB function has been used for most of this work. 

\begin{figure}[!t]
    \centering
    \includegraphics[width=0.46\textwidth]{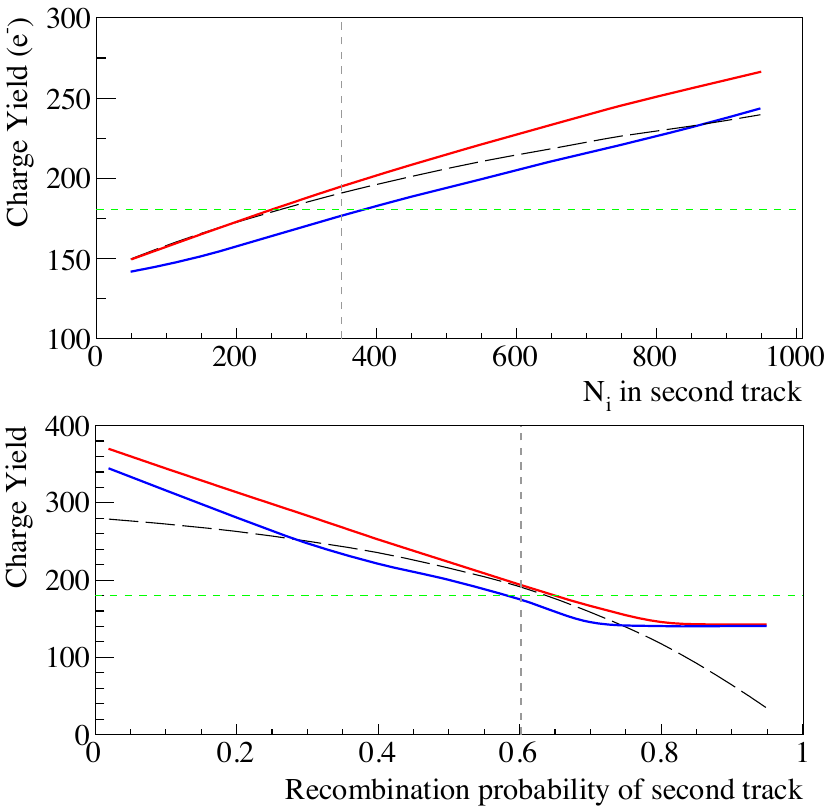}
    \caption{The charge yield of hypothetical $^{124}$Xe DEC-like events where one constituent track has a varying ion count (top) or recombination probability (bottom). Results from the constant-$r$ (blue) and TIB-based running-$r$ (red) methods are compared to that from simply extrapolating the TIB model by adding $\xi$'s (black, dashed). The vertical gray dotted lines indicate the physical values of $N_i$ ($\sim$5 keV per track, $\sim$10 keV total) and $\bar{r}$; the horizontal green dotted lines indicate the LZ best fit. }
    \label{fig:testtib}
\end{figure}

For recombination in composite events consisting of similar tracks, such as in a $^{124}$Xe DEC decay, the TIB model can be used by scaling up the number of total ions in the event from the single-track value. 
Within the TIB framework, this scaling represents an increase in the ionization density without changing the position distribution. 
As an approximation, the TIB model can be generalized by adding the $\xi$ contribution of each track, $\xi_k$, as defined 
in Eq.~\ref{eq:tib} to allow slight variations in the properties of the constituent tracks: 
\begin{equation}
    \bar{r}=1-\frac{ln(1+\xi)}{\xi}, \; \xi=\sum_k \xi_k \label{eq:tibext}
\end{equation}

Charge yields estimated using this TIB extrapolation for both $^{124}$Xe DEC and Migdal signals are found to be similar to those using the TIB-based running-$r$ method. This TIB extrapolation method is also studied in Ref.~\cite{LZ2025_EC}, which reports similar results. 
To further study the relationship of the two TIB-based approaches, we compare the predicted charge yields for hypothetical $^{124}$Xe DEC-like events where the ion count (Fig.~\ref{fig:testtib}, top) or the recombination probability (Fig.~\ref{fig:testtib}, bottom) of one constituent track is allowed to vary. 
When the ion count is varied, in the physical region near the single-L EC energy, the results from different models (red: TIB-based running-$r$, blue: constant-$r$, dashed black: TIB extrapolation in Eq.~\ref{eq:tibext}) closely track each other. 
However, at high ion counts Eq.~\ref{eq:tibext} predicts a charge yield that is lower than the constant-$r$ yield prediction, which is interpreted as the lower bound in our model. 
This behavior is consistent with a known limitation of TIB for overestimating recombination above 10 keV beta energy, which is mitigated in \cite{DahlThesis} by incorporating the growth of the track size.  
In Fig.~\ref{fig:testtib} (bottom), excellent agreement is observed near the physical recombination probability of $\bar{r}$=0.6, but the TIB extrapolation predicts significantly lower charge yield in unphysical regions. 
Most notably, when one track is assigned $\bar{r}\sim$100\%, Eq.~\ref{eq:tibext} predicts $\sim$100\% recombination for the whole event and causes all electrons and ions to recombine. 
However, in our model, a hypothetical track with 100\% recombination probability ceases to exist after all its ions recombine with all of its electrons, and would have no real effect on the charge yield of other tracks. 
Similarly a hypothetical track with 0\% recombination probability only contributes electrons and its ability to recombine with electrons will not be affected by a colocated track. 
Therefore, we conclude that, when TIB is used as an input, the running-$r$ model developed in this work is compatible with the TIB extrapolation where TIB is valid, but our model retains prediction power beyond the applicable scope of TIB.  

\section{Conclusion}
\label{sec:concl}

We have developed a data-driven model to evaluate the electron-ion recombination strength in composite interactions, where multiple ionizing particles produce simultaneous and overlapping ionization signatures. These events represent a significant source of uncertainty in both dark matter search experiments and studies of the atomic Migdal effect.
Our model, using single-track interaction data as input, generates predictions that strongly constrain the charge yields of composite interactions in liquid xenon, 
and the results are in good agreement with available data. 
While the method depends on the assumed exciton-to-ion ratio from particle interactions in liquid xenon, this dependence has been verified to be minimal, rendering the model robust. 
The method is expected to remain valid for both electron recoils and nuclear recoils relevant to dark matter search experiments.

\begin{acknowledgments}

 This project is supported by the U.S. Department of Energy (DOE) Office of Science, 
 Office of High Energy Physics under Work Proposal Number SCW1676 awarded to Lawrence Livermore National Laboratory (LLNL). 
B.~L. was supported by the DOE Laboratory Directed Research and Development program at SLAC National Accelerator Laboratory, under contract DE-AC02-76F00515 as part of the Panofsky Fellowship. 
C.~L, M.~S and G.~B are supported by the DOE award DE-SC0015535 to the University at Albany - State University of New York. 

This work was performed under the auspices of the U.S. Department of Energy by Lawrence Livermore
National Laboratory under Contract DE-AC52-07NA27344. 
LLNL release number: LLNL-JRNL-2002846. 
\end{acknowledgments}

\bibliographystyle{apsrev}
\bibliography{biblio}

\begin{thebibliography}{43}
\expandafter\ifx\csname natexlab\endcsname\relax\def\natexlab#1{#1}\fi
\expandafter\ifx\csname bibnamefont\endcsname\relax
  \def\bibnamefont#1{#1}\fi
\expandafter\ifx\csname bibfnamefont\endcsname\relax
  \def\bibfnamefont#1{#1}\fi
\expandafter\ifx\csname citenamefont\endcsname\relax
  \def\citenamefont#1{#1}\fi
\expandafter\ifx\csname url\endcsname\relax
  \def\url#1{\texttt{#1}}\fi
\expandafter\ifx\csname urlprefix\endcsname\relax\def\urlprefix{URL }\fi
\providecommand{\bibinfo}[2]{#2}
\providecommand{\eprint}[2][]{\url{#2}}

\bibitem[{\citenamefont{Aprile et~al.}(2006)\citenamefont{Aprile, Dahl,
  de~Viveiros, Gaitskell, Giboni, Kwong, Majewski, Ni, Shutt, and
  Yamashita}}]{Xenon10_ERNR}
\bibinfo{author}{\bibfnamefont{E.}~\bibnamefont{Aprile}},
  \bibinfo{author}{\bibfnamefont{C.~E.} \bibnamefont{Dahl}},
  \bibinfo{author}{\bibfnamefont{L.}~\bibnamefont{de~Viveiros}},
  \bibinfo{author}{\bibfnamefont{R.~J.} \bibnamefont{Gaitskell}},
  \bibinfo{author}{\bibfnamefont{K.~L.} \bibnamefont{Giboni}},
  \bibinfo{author}{\bibfnamefont{J.}~\bibnamefont{Kwong}},
  \bibinfo{author}{\bibfnamefont{P.}~\bibnamefont{Majewski}},
  \bibinfo{author}{\bibfnamefont{K.}~\bibnamefont{Ni}},
  \bibinfo{author}{\bibfnamefont{T.}~\bibnamefont{Shutt}}, \bibnamefont{and}
  \bibinfo{author}{\bibfnamefont{M.}~\bibnamefont{Yamashita}},
  \href{http://dx.doi.org/10.1103/PhysRevLett.97.081302}{\bibinfo{journal}{Phys.
  Rev. Lett.}, \textbf{\bibinfo{volume}{97}},
  \bibinfo{pages}{081302}\bibinfo{year}{ (\bibinfo{year}{2006})}}.

\bibitem[{\citenamefont{Aalbers
  et~al.}(2024{\natexlab{a}})\citenamefont{Aalbers, Akerib, Musalhi, Alder,
  Amarasinghe, Ames, Anderson, Angelides, Araújo, Armstrong
  et~al.}}]{LZ2024_WS}
\bibinfo{author}{\bibfnamefont{J.}~\bibnamefont{Aalbers}},
  \bibinfo{author}{\bibfnamefont{D.~S.} \bibnamefont{Akerib}},
  \bibinfo{author}{\bibfnamefont{A.~K.~A.} \bibnamefont{Musalhi}},
  \bibinfo{author}{\bibfnamefont{F.}~\bibnamefont{Alder}},
  \bibinfo{author}{\bibfnamefont{C.~S.} \bibnamefont{Amarasinghe}},
  \bibinfo{author}{\bibfnamefont{A.}~\bibnamefont{Ames}},
  \bibinfo{author}{\bibfnamefont{T.~J.} \bibnamefont{Anderson}},
  \bibinfo{author}{\bibfnamefont{N.}~\bibnamefont{Angelides}},
  \bibinfo{author}{\bibfnamefont{H.~M.} \bibnamefont{Araújo}},
  \bibinfo{author}{\bibfnamefont{J.~E.} \bibnamefont{Armstrong}},
  \bibnamefont{et~al.} (\bibinfo{collaboration}{LZ Collaboration}),
  \emph{\bibinfo{title}{Dark matter search results from 4.2 tonne-years of
  exposure of the lux-zeplin (lz) experiment}}
  (\bibinfo{year}{2024}{\natexlab{a}}), \eprint{2410.17036},
  \urlprefix\url{https://arxiv.org/abs/2410.17036}.

\bibitem[{\citenamefont{Aprile et~al.}(2024)\citenamefont{Aprile, Aalbers, Abe,
  Ahmed~Maouloud, Althueser, Andrieu, Angelino, Ant\'on~Martin, Arneodo, Baudis
  et~al.}}]{XENON2024_8B}
\bibinfo{author}{\bibfnamefont{E.}~\bibnamefont{Aprile}},
  \bibinfo{author}{\bibfnamefont{J.}~\bibnamefont{Aalbers}},
  \bibinfo{author}{\bibfnamefont{K.}~\bibnamefont{Abe}},
  \bibinfo{author}{\bibfnamefont{S.}~\bibnamefont{Ahmed~Maouloud}},
  \bibinfo{author}{\bibfnamefont{L.}~\bibnamefont{Althueser}},
  \bibinfo{author}{\bibfnamefont{B.}~\bibnamefont{Andrieu}},
  \bibinfo{author}{\bibfnamefont{E.}~\bibnamefont{Angelino}},
  \bibinfo{author}{\bibfnamefont{D.}~\bibnamefont{Ant\'on~Martin}},
  \bibinfo{author}{\bibfnamefont{F.}~\bibnamefont{Arneodo}},
  \bibinfo{author}{\bibfnamefont{L.}~\bibnamefont{Baudis}},
  \bibnamefont{et~al.} (\bibinfo{collaboration}{XENON Collaboration}),
  \href{http://dx.doi.org/10.1103/PhysRevLett.133.191002}{\bibinfo{journal}{Phys.
  Rev. Lett.}, \textbf{\bibinfo{volume}{133}},
  \bibinfo{pages}{191002}\bibinfo{year}{ (\bibinfo{year}{2024})}}.

\bibitem[{\citenamefont{Bo et~al.}(2024)\citenamefont{Bo, Chen, Chen, Chen,
  Cheng, Cui, Fan, Fang, Gao, Geng et~al.}}]{PandaX2024_8B}
\bibinfo{author}{\bibfnamefont{Z.}~\bibnamefont{Bo}},
  \bibinfo{author}{\bibfnamefont{W.}~\bibnamefont{Chen}},
  \bibinfo{author}{\bibfnamefont{X.}~\bibnamefont{Chen}},
  \bibinfo{author}{\bibfnamefont{Y.}~\bibnamefont{Chen}},
  \bibinfo{author}{\bibfnamefont{Z.}~\bibnamefont{Cheng}},
  \bibinfo{author}{\bibfnamefont{X.}~\bibnamefont{Cui}},
  \bibinfo{author}{\bibfnamefont{Y.}~\bibnamefont{Fan}},
  \bibinfo{author}{\bibfnamefont{D.}~\bibnamefont{Fang}},
  \bibinfo{author}{\bibfnamefont{Z.}~\bibnamefont{Gao}},
  \bibinfo{author}{\bibfnamefont{L.}~\bibnamefont{Geng}}, \bibnamefont{et~al.}
  (\bibinfo{collaboration}{PandaX Collaboration}),
  \href{http://dx.doi.org/10.1103/PhysRevLett.133.191001}{\bibinfo{journal}{Phys.
  Rev. Lett.}, \textbf{\bibinfo{volume}{133}},
  \bibinfo{pages}{191001}\bibinfo{year}{ (\bibinfo{year}{2024})}}.

\bibitem[{\citenamefont{Adhikari et~al.}(2021)\citenamefont{Adhikari,
  Al~Kharusi, Angelico, Anton, Arnquist, Badhrees, Bane, Belov, Bernard, Bhatta
  et~al.}}]{adhikari2021nexo}
\bibinfo{author}{\bibfnamefont{G.}~\bibnamefont{Adhikari}},
  \bibinfo{author}{\bibfnamefont{S.}~\bibnamefont{Al~Kharusi}},
  \bibinfo{author}{\bibfnamefont{E.}~\bibnamefont{Angelico}},
  \bibinfo{author}{\bibfnamefont{G.}~\bibnamefont{Anton}},
  \bibinfo{author}{\bibfnamefont{I.}~\bibnamefont{Arnquist}},
  \bibinfo{author}{\bibfnamefont{I.}~\bibnamefont{Badhrees}},
  \bibinfo{author}{\bibfnamefont{J.}~\bibnamefont{Bane}},
  \bibinfo{author}{\bibfnamefont{V.}~\bibnamefont{Belov}},
  \bibinfo{author}{\bibfnamefont{E.}~\bibnamefont{Bernard}},
  \bibinfo{author}{\bibfnamefont{T.}~\bibnamefont{Bhatta}},
  \bibnamefont{et~al.},
  \href{http://dx.doi.org/10.1088/1361-6471/ac3631}{\bibinfo{journal}{Journal
  of Physics G: Nuclear and Particle Physics}, \textbf{\bibinfo{volume}{49}},
  \bibinfo{pages}{015104}\bibinfo{year}{ (\bibinfo{year}{2021})}}.

\bibitem[{\citenamefont{Dahl}(2009)}]{DahlThesis}
\bibinfo{author}{\bibfnamefont{C.~E.} \bibnamefont{Dahl}}, Ph.D. thesis,
  \bibinfo{school}{Princeton U.} (\bibinfo{year}{2009}),
  \urlprefix\url{https://www.princeton.edu/physics/graduate-program/theses/theses-from-2009/E.Dahlthesis.pdf}.

\bibitem[{\citenamefont{Szydagis et~al.}(2025)\citenamefont{Szydagis, Balajthy,
  Block, Brodsky, Brown, Cutter, Farrell, Huang, Kamaha, Kozlova
  et~al.}}]{Szydagis2025_NEST}
\bibinfo{author}{\bibfnamefont{M.}~\bibnamefont{Szydagis}},
  \bibinfo{author}{\bibfnamefont{J.}~\bibnamefont{Balajthy}},
  \bibinfo{author}{\bibfnamefont{G.~A.} \bibnamefont{Block}},
  \bibinfo{author}{\bibfnamefont{J.~P.} \bibnamefont{Brodsky}},
  \bibinfo{author}{\bibfnamefont{E.}~\bibnamefont{Brown}},
  \bibinfo{author}{\bibfnamefont{J.~E.} \bibnamefont{Cutter}},
  \bibinfo{author}{\bibfnamefont{S.~J.} \bibnamefont{Farrell}},
  \bibinfo{author}{\bibfnamefont{J.}~\bibnamefont{Huang}},
  \bibinfo{author}{\bibfnamefont{A.~C.} \bibnamefont{Kamaha}},
  \bibinfo{author}{\bibfnamefont{E.~S.} \bibnamefont{Kozlova}},
  \bibnamefont{et~al.},
  \href{http://dx.doi.org/10.3389/fdest.2024.1480975}{\bibinfo{journal}{Frontiers
  in Detector Science and Technology},
  \textbf{\bibinfo{volume}{2}}\bibinfo{year}{ (\bibinfo{year}{2025})}}, ISSN
  \bibinfo{issn}{2813-8031}.

\bibitem[{\citenamefont{Lenardo et~al.}(2019)\citenamefont{Lenardo, Xu,
  Pereverzev, Akindele, Naim, Kingston, Bernstein, Kazkaz, Tripathi, Awe
  et~al.}}]{Lenardo2019_XeNR}
\bibinfo{author}{\bibfnamefont{B.~G.} \bibnamefont{Lenardo}},
  \bibinfo{author}{\bibfnamefont{J.}~\bibnamefont{Xu}},
  \bibinfo{author}{\bibfnamefont{S.}~\bibnamefont{Pereverzev}},
  \bibinfo{author}{\bibfnamefont{O.~A.} \bibnamefont{Akindele}},
  \bibinfo{author}{\bibfnamefont{D.}~\bibnamefont{Naim}},
  \bibinfo{author}{\bibfnamefont{J.}~\bibnamefont{Kingston}},
  \bibinfo{author}{\bibfnamefont{A.}~\bibnamefont{Bernstein}},
  \bibinfo{author}{\bibfnamefont{K.}~\bibnamefont{Kazkaz}},
  \bibinfo{author}{\bibfnamefont{M.}~\bibnamefont{Tripathi}},
  \bibinfo{author}{\bibfnamefont{C.}~\bibnamefont{Awe}}, \bibnamefont{et~al.},
  \href{http://dx.doi.org/10.1103/PhysRevLett.123.231106}{\bibinfo{journal}{Phys.
  Rev. Lett.}, \textbf{\bibinfo{volume}{123}},
  \bibinfo{pages}{231106}\bibinfo{year}{ (\bibinfo{year}{2019})}}.

\bibitem[{\citenamefont{Akerib et~al.}(2025)\citenamefont{Akerib, Alsum,
  Ara\'ujo, Bai, Balajthy, Bang, Baxter, Bernard, Bernstein, Biesiadzinski
  et~al.}}]{LUX2025_NR}
\bibinfo{author}{\bibfnamefont{D.~S.} \bibnamefont{Akerib}},
  \bibinfo{author}{\bibfnamefont{S.}~\bibnamefont{Alsum}},
  \bibinfo{author}{\bibfnamefont{H.~M.} \bibnamefont{Ara\'ujo}},
  \bibinfo{author}{\bibfnamefont{X.}~\bibnamefont{Bai}},
  \bibinfo{author}{\bibfnamefont{J.}~\bibnamefont{Balajthy}},
  \bibinfo{author}{\bibfnamefont{J.}~\bibnamefont{Bang}},
  \bibinfo{author}{\bibfnamefont{A.}~\bibnamefont{Baxter}},
  \bibinfo{author}{\bibfnamefont{E.~P.} \bibnamefont{Bernard}},
  \bibinfo{author}{\bibfnamefont{A.}~\bibnamefont{Bernstein}},
  \bibinfo{author}{\bibfnamefont{T.~P.} \bibnamefont{Biesiadzinski}},
  \bibnamefont{et~al.},
  \href{http://dx.doi.org/10.1103/PhysRevLett.134.061002}{\bibinfo{journal}{Phys.
  Rev. Lett.}, \textbf{\bibinfo{volume}{134}},
  \bibinfo{pages}{061002}\bibinfo{year}{ (\bibinfo{year}{2025})}}.

\bibitem[{\citenamefont{Pershing et~al.}(2022)\citenamefont{Pershing, Naim,
  Lenardo, Xu, Kingston, Mizrachi, Mozin, Kerr, Pereverzev, Bernstein
  et~al.}}]{Pershing2022_XeNR}
\bibinfo{author}{\bibfnamefont{T.}~\bibnamefont{Pershing}},
  \bibinfo{author}{\bibfnamefont{D.}~\bibnamefont{Naim}},
  \bibinfo{author}{\bibfnamefont{B.~G.} \bibnamefont{Lenardo}},
  \bibinfo{author}{\bibfnamefont{J.}~\bibnamefont{Xu}},
  \bibinfo{author}{\bibfnamefont{J.}~\bibnamefont{Kingston}},
  \bibinfo{author}{\bibfnamefont{E.}~\bibnamefont{Mizrachi}},
  \bibinfo{author}{\bibfnamefont{V.}~\bibnamefont{Mozin}},
  \bibinfo{author}{\bibfnamefont{P.}~\bibnamefont{Kerr}},
  \bibinfo{author}{\bibfnamefont{S.}~\bibnamefont{Pereverzev}},
  \bibinfo{author}{\bibfnamefont{A.}~\bibnamefont{Bernstein}},
  \bibnamefont{et~al.},
  \href{http://dx.doi.org/10.1103/PhysRevD.106.052013}{\bibinfo{journal}{Phys.
  Rev. D}, \textbf{\bibinfo{volume}{106}},
  \bibinfo{pages}{052013}\bibinfo{year}{ (\bibinfo{year}{2022})}}.

\bibitem[{\citenamefont{{XENON Collaboration}}(2023)}]{XENON2023_Ar37}
\bibinfo{author}{\bibnamefont{{XENON Collaboration}}},
  \href{http://dx.doi.org/10.1140/epjc/s10052-023-11512-z}{\bibinfo{journal}{The
  European Physical Journal C}, \textbf{\bibinfo{volume}{83}},
  \bibinfo{pages}{542}\bibinfo{year}{ (\bibinfo{year}{2023})}}.

\bibitem[{\citenamefont{Akerib et~al.}(2016)\citenamefont{Akerib, Ara\'ujo,
  Bai, Bailey, Balajthy, Beltrame, Bernard, Bernstein, Biesiadzinski, Boulton
  et~al.}}]{LUX2016_3H}
\bibinfo{author}{\bibfnamefont{D.~S.} \bibnamefont{Akerib}},
  \bibinfo{author}{\bibfnamefont{H.~M.} \bibnamefont{Ara\'ujo}},
  \bibinfo{author}{\bibfnamefont{X.}~\bibnamefont{Bai}},
  \bibinfo{author}{\bibfnamefont{A.~J.} \bibnamefont{Bailey}},
  \bibinfo{author}{\bibfnamefont{J.}~\bibnamefont{Balajthy}},
  \bibinfo{author}{\bibfnamefont{P.}~\bibnamefont{Beltrame}},
  \bibinfo{author}{\bibfnamefont{E.~P.} \bibnamefont{Bernard}},
  \bibinfo{author}{\bibfnamefont{A.}~\bibnamefont{Bernstein}},
  \bibinfo{author}{\bibfnamefont{T.~P.} \bibnamefont{Biesiadzinski}},
  \bibinfo{author}{\bibfnamefont{E.~M.} \bibnamefont{Boulton}},
  \bibnamefont{et~al.} (\bibinfo{collaboration}{LUX Collaboration}),
  \href{http://dx.doi.org/10.1103/PhysRevD.93.072009}{\bibinfo{journal}{Phys.
  Rev. D}, \textbf{\bibinfo{volume}{93}},
  \bibinfo{pages}{072009}\bibinfo{year}{ (\bibinfo{year}{2016})}}.

\bibitem[{\citenamefont{Boulton et~al.}(2017)\citenamefont{Boulton, Bernard,
  Destefano, Edwards, Gai, Hertel, Horn, Larsen, Tennyson, Wahl
  et~al.}}]{Boulton2017_Ar37}
\bibinfo{author}{\bibfnamefont{E.}~\bibnamefont{Boulton}},
  \bibinfo{author}{\bibfnamefont{E.}~\bibnamefont{Bernard}},
  \bibinfo{author}{\bibfnamefont{N.}~\bibnamefont{Destefano}},
  \bibinfo{author}{\bibfnamefont{B.}~\bibnamefont{Edwards}},
  \bibinfo{author}{\bibfnamefont{M.}~\bibnamefont{Gai}},
  \bibinfo{author}{\bibfnamefont{S.}~\bibnamefont{Hertel}},
  \bibinfo{author}{\bibfnamefont{M.}~\bibnamefont{Horn}},
  \bibinfo{author}{\bibfnamefont{N.}~\bibnamefont{Larsen}},
  \bibinfo{author}{\bibfnamefont{B.}~\bibnamefont{Tennyson}},
  \bibinfo{author}{\bibfnamefont{C.}~\bibnamefont{Wahl}}, \bibnamefont{et~al.},
   \href{http://stacks.iop.org/1748-0221/12/i=08/a=P08004}{\bibinfo{journal}{Journal
  of Instrumentation}, \textbf{\bibinfo{volume}{12}},
  \bibinfo{pages}{P08004}\bibinfo{year}{ (\bibinfo{year}{2017})}}.

\bibitem[{\citenamefont{Akerib et~al.}(2017{\natexlab{a}})\citenamefont{Akerib,
  Alsum, Ara\'ujo, Bai, Bailey, Balajthy, Beltrame, Bernard, Bernstein,
  Biesiadzinski et~al.}}]{LUX2017_Kr83m}
\bibinfo{author}{\bibfnamefont{D.~S.} \bibnamefont{Akerib}},
  \bibinfo{author}{\bibfnamefont{S.}~\bibnamefont{Alsum}},
  \bibinfo{author}{\bibfnamefont{H.~M.} \bibnamefont{Ara\'ujo}},
  \bibinfo{author}{\bibfnamefont{X.}~\bibnamefont{Bai}},
  \bibinfo{author}{\bibfnamefont{A.~J.} \bibnamefont{Bailey}},
  \bibinfo{author}{\bibfnamefont{J.}~\bibnamefont{Balajthy}},
  \bibinfo{author}{\bibfnamefont{P.}~\bibnamefont{Beltrame}},
  \bibinfo{author}{\bibfnamefont{E.~P.} \bibnamefont{Bernard}},
  \bibinfo{author}{\bibfnamefont{A.}~\bibnamefont{Bernstein}},
  \bibinfo{author}{\bibfnamefont{T.~P.} \bibnamefont{Biesiadzinski}},
  \bibnamefont{et~al.} (\bibinfo{collaboration}{LUX Collaboration}),
  \href{http://dx.doi.org/10.1103/PhysRevD.96.112009}{\bibinfo{journal}{Phys.
  Rev. D}, \textbf{\bibinfo{volume}{96}},
  \bibinfo{pages}{112009}\bibinfo{year}{
  (\bibinfo{year}{2017}{\natexlab{a}})}}.

\bibitem[{\citenamefont{Akerib et~al.}(2017{\natexlab{b}})\citenamefont{Akerib,
  Alsum, Ara\'ujo, Bai, Bailey, Balajthy, Beltrame, Bernard, Bernstein,
  Biesiadzinski et~al.}}]{LUX2017_Xe127}
\bibinfo{author}{\bibfnamefont{D.~S.} \bibnamefont{Akerib}},
  \bibinfo{author}{\bibfnamefont{S.}~\bibnamefont{Alsum}},
  \bibinfo{author}{\bibfnamefont{H.~M.} \bibnamefont{Ara\'ujo}},
  \bibinfo{author}{\bibfnamefont{X.}~\bibnamefont{Bai}},
  \bibinfo{author}{\bibfnamefont{A.~J.} \bibnamefont{Bailey}},
  \bibinfo{author}{\bibfnamefont{J.}~\bibnamefont{Balajthy}},
  \bibinfo{author}{\bibfnamefont{P.}~\bibnamefont{Beltrame}},
  \bibinfo{author}{\bibfnamefont{E.~P.} \bibnamefont{Bernard}},
  \bibinfo{author}{\bibfnamefont{A.}~\bibnamefont{Bernstein}},
  \bibinfo{author}{\bibfnamefont{T.~P.} \bibnamefont{Biesiadzinski}},
  \bibnamefont{et~al.},
  \href{http://dx.doi.org/10.1103/PhysRevD.96.112011}{\bibinfo{journal}{Phys.
  Rev. D}, \textbf{\bibinfo{volume}{96}},
  \bibinfo{pages}{112011}\bibinfo{year}{
  (\bibinfo{year}{2017}{\natexlab{b}})}}.

\bibitem[{\citenamefont{Temples et~al.}(2021)\citenamefont{Temples, McLaughlin,
  Bargemann, Baxter, Cottle, Dahl, Lippincott, Monte, and
  Phelan}}]{Temples2021_127Xe}
\bibinfo{author}{\bibfnamefont{D.~J.} \bibnamefont{Temples}},
  \bibinfo{author}{\bibfnamefont{J.}~\bibnamefont{McLaughlin}},
  \bibinfo{author}{\bibfnamefont{J.}~\bibnamefont{Bargemann}},
  \bibinfo{author}{\bibfnamefont{D.}~\bibnamefont{Baxter}},
  \bibinfo{author}{\bibfnamefont{A.}~\bibnamefont{Cottle}},
  \bibinfo{author}{\bibfnamefont{C.~E.} \bibnamefont{Dahl}},
  \bibinfo{author}{\bibfnamefont{W.~H.} \bibnamefont{Lippincott}},
  \bibinfo{author}{\bibfnamefont{A.}~\bibnamefont{Monte}}, \bibnamefont{and}
  \bibinfo{author}{\bibfnamefont{J.}~\bibnamefont{Phelan}},
  \href{http://dx.doi.org/10.1103/PhysRevD.104.112001}{\bibinfo{journal}{Phys.
  Rev. D}, \textbf{\bibinfo{volume}{104}},
  \bibinfo{pages}{112001}\bibinfo{year}{ (\bibinfo{year}{2021})}}.

\bibitem[{\citenamefont{Aprile et~al.}(2017)\citenamefont{Aprile, Aalbers,
  Agostini, Alfonsi, Amaro, Anthony, Arneodo, Barrow, Baudis, Bauermeister
  et~al.}}]{XENON1T_2017}
\bibinfo{author}{\bibfnamefont{E.}~\bibnamefont{Aprile}},
  \bibinfo{author}{\bibfnamefont{J.}~\bibnamefont{Aalbers}},
  \bibinfo{author}{\bibfnamefont{F.}~\bibnamefont{Agostini}},
  \bibinfo{author}{\bibfnamefont{M.}~\bibnamefont{Alfonsi}},
  \bibinfo{author}{\bibfnamefont{F.~D.} \bibnamefont{Amaro}},
  \bibinfo{author}{\bibfnamefont{M.}~\bibnamefont{Anthony}},
  \bibinfo{author}{\bibfnamefont{F.}~\bibnamefont{Arneodo}},
  \bibinfo{author}{\bibfnamefont{P.}~\bibnamefont{Barrow}},
  \bibinfo{author}{\bibfnamefont{L.}~\bibnamefont{Baudis}},
  \bibinfo{author}{\bibfnamefont{B.}~\bibnamefont{Bauermeister}},
  \bibnamefont{et~al.} (\bibinfo{collaboration}{XENON Collaboration}),
  \href{http://dx.doi.org/10.1103/PhysRevLett.119.181301}{\bibinfo{journal}{Phys.
  Rev. Lett.}, \textbf{\bibinfo{volume}{119}},
  \bibinfo{pages}{181301}\bibinfo{year}{ (\bibinfo{year}{2017})}}.

\bibitem[{\citenamefont{Akimov et~al.}(2014)\citenamefont{Akimov, Afanasyev,
  Alexandrov, Belov, Bolozdynya, Burenkov, Efremenko, Egorov, Etenko, Gulin
  et~al.}}]{Akimov2014_Ar37}
\bibinfo{author}{\bibfnamefont{D.}~\bibnamefont{Akimov}},
  \bibinfo{author}{\bibfnamefont{V.}~\bibnamefont{Afanasyev}},
  \bibinfo{author}{\bibfnamefont{I.}~\bibnamefont{Alexandrov}},
  \bibinfo{author}{\bibfnamefont{V.}~\bibnamefont{Belov}},
  \bibinfo{author}{\bibfnamefont{A.}~\bibnamefont{Bolozdynya}},
  \bibinfo{author}{\bibfnamefont{A.}~\bibnamefont{Burenkov}},
  \bibinfo{author}{\bibfnamefont{Y.}~\bibnamefont{Efremenko}},
  \bibinfo{author}{\bibfnamefont{D.}~\bibnamefont{Egorov}},
  \bibinfo{author}{\bibfnamefont{A.}~\bibnamefont{Etenko}},
  \bibinfo{author}{\bibfnamefont{M.}~\bibnamefont{Gulin}},
  \bibnamefont{et~al.},
  \href{http://stacks.iop.org/1748-0221/9/i=11/a=P11014}{\bibinfo{journal}{Journal
  of Instrumentation}, \textbf{\bibinfo{volume}{9}},
  \bibinfo{pages}{P11014}\bibinfo{year}{ (\bibinfo{year}{2014})}}.

\bibitem[{\citenamefont{Aalbers et~al.}(2025)\citenamefont{Aalbers, Akerib,
  Al~Musalhi, Alder, Amarasinghe, Ames, Anderson, Angelides, Ara{\'u}jo,
  Armstrong et~al.}}]{LZ2025_EC}
\bibinfo{author}{\bibfnamefont{J.}~\bibnamefont{Aalbers}},
  \bibinfo{author}{\bibfnamefont{D.}~\bibnamefont{Akerib}},
  \bibinfo{author}{\bibfnamefont{A.}~\bibnamefont{Al~Musalhi}},
  \bibinfo{author}{\bibfnamefont{F.}~\bibnamefont{Alder}},
  \bibinfo{author}{\bibfnamefont{C.}~\bibnamefont{Amarasinghe}},
  \bibinfo{author}{\bibfnamefont{A.}~\bibnamefont{Ames}},
  \bibinfo{author}{\bibfnamefont{T.}~\bibnamefont{Anderson}},
  \bibinfo{author}{\bibfnamefont{N.}~\bibnamefont{Angelides}},
  \bibinfo{author}{\bibfnamefont{H.}~\bibnamefont{Ara{\'u}jo}},
  \bibinfo{author}{\bibfnamefont{J.}~\bibnamefont{Armstrong}},
  \bibnamefont{et~al.} (\bibinfo{collaboration}{LZ Collaboration}),
  \emph{\bibinfo{title}{Measurements and models of enhanced recombination
  following inner-shell vacancies in liquid xenon}} (\bibinfo{year}{2025}),
  \eprint{2502.X}, \urlprefix\url{https://arxiv.org/abs/2502.X}.

\bibitem[{\citenamefont{Aprile et~al.}(2019{\natexlab{a}})\citenamefont{Aprile,
  Aalbers, Agostini, Alfonsi, Althueser, Amaro, Antochi, Angelino, Arneodo,
  Barge et~al.}}]{Xenon2019_124Xe}
\bibinfo{author}{\bibfnamefont{E.}~\bibnamefont{Aprile}},
  \bibinfo{author}{\bibfnamefont{J.}~\bibnamefont{Aalbers}},
  \bibinfo{author}{\bibfnamefont{F.}~\bibnamefont{Agostini}},
  \bibinfo{author}{\bibfnamefont{M.}~\bibnamefont{Alfonsi}},
  \bibinfo{author}{\bibfnamefont{L.}~\bibnamefont{Althueser}},
  \bibinfo{author}{\bibfnamefont{F.~D.} \bibnamefont{Amaro}},
  \bibinfo{author}{\bibfnamefont{V.~C.} \bibnamefont{Antochi}},
  \bibinfo{author}{\bibfnamefont{E.}~\bibnamefont{Angelino}},
  \bibinfo{author}{\bibfnamefont{F.}~\bibnamefont{Arneodo}},
  \bibinfo{author}{\bibfnamefont{D.}~\bibnamefont{Barge}}, \bibnamefont{et~al.}
  (\bibinfo{collaboration}{{XENON Collaboration}}),
  \href{http://dx.doi.org/https://doi.org/10.1038/s41586-019-1124-4}{\bibinfo{journal}{Nature},
  \textbf{\bibinfo{volume}{568}}, \bibinfo{pages}{532}\bibinfo{year}{
  (\bibinfo{year}{2019}{\natexlab{a}})}}.

\bibitem[{\citenamefont{Aalbers
  et~al.}(2024{\natexlab{b}})\citenamefont{Aalbers, Akerib, Al~Musalhi, Alder,
  Amarasinghe, Ames, Anderson, Angelides, Ara{\'u}jo, Armstrong
  et~al.}}]{LZ2024_124Xe}
\bibinfo{author}{\bibfnamefont{J.}~\bibnamefont{Aalbers}},
  \bibinfo{author}{\bibfnamefont{D.}~\bibnamefont{Akerib}},
  \bibinfo{author}{\bibfnamefont{A.}~\bibnamefont{Al~Musalhi}},
  \bibinfo{author}{\bibfnamefont{F.}~\bibnamefont{Alder}},
  \bibinfo{author}{\bibfnamefont{C.}~\bibnamefont{Amarasinghe}},
  \bibinfo{author}{\bibfnamefont{A.}~\bibnamefont{Ames}},
  \bibinfo{author}{\bibfnamefont{T.}~\bibnamefont{Anderson}},
  \bibinfo{author}{\bibfnamefont{N.}~\bibnamefont{Angelides}},
  \bibinfo{author}{\bibfnamefont{H.}~\bibnamefont{Ara{\'u}jo}},
  \bibinfo{author}{\bibfnamefont{J.}~\bibnamefont{Armstrong}},
  \bibnamefont{et~al.} (\bibinfo{collaboration}{LZ Collaboration}),
  \href{http://dx.doi.org/10.1088/1361-6471/ad9039}{\bibinfo{journal}{Journal
  of Physics G: Nuclear and Particle Physics}, \textbf{\bibinfo{volume}{52}},
  \bibinfo{pages}{015103}\bibinfo{year}{
  (\bibinfo{year}{2024}{\natexlab{b}})}}.

\bibitem[{\citenamefont{Aprile et~al.}(2022)\citenamefont{Aprile, Abe,
  Agostini, Ahmed~Maouloud, Althueser, Andrieu, Angelino, Angevaare, Antochi,
  Ant\'on~Martin et~al.}}]{XnT2022_NewPhysics}
\bibinfo{author}{\bibfnamefont{E.}~\bibnamefont{Aprile}},
  \bibinfo{author}{\bibfnamefont{K.}~\bibnamefont{Abe}},
  \bibinfo{author}{\bibfnamefont{F.}~\bibnamefont{Agostini}},
  \bibinfo{author}{\bibfnamefont{S.}~\bibnamefont{Ahmed~Maouloud}},
  \bibinfo{author}{\bibfnamefont{L.}~\bibnamefont{Althueser}},
  \bibinfo{author}{\bibfnamefont{B.}~\bibnamefont{Andrieu}},
  \bibinfo{author}{\bibfnamefont{E.}~\bibnamefont{Angelino}},
  \bibinfo{author}{\bibfnamefont{J.~R.} \bibnamefont{Angevaare}},
  \bibinfo{author}{\bibfnamefont{V.~C.} \bibnamefont{Antochi}},
  \bibinfo{author}{\bibfnamefont{D.}~\bibnamefont{Ant\'on~Martin}},
  \bibnamefont{et~al.} (\bibinfo{collaboration}{XENON Collaboration}),
  \href{http://dx.doi.org/10.1103/PhysRevLett.129.161805}{\bibinfo{journal}{Phys.
  Rev. Lett.}, \textbf{\bibinfo{volume}{129}},
  \bibinfo{pages}{161805}\bibinfo{year}{ (\bibinfo{year}{2022})}}.

\bibitem[{\citenamefont{Bo et~al.}(2025)\citenamefont{Bo, Chen, Chen, Chen,
  Cheng, Cui, Fan, Fang, Gao, Geng et~al.}}]{Panda2024_124Xe}
\bibinfo{author}{\bibfnamefont{Z.}~\bibnamefont{Bo}},
  \bibinfo{author}{\bibfnamefont{W.}~\bibnamefont{Chen}},
  \bibinfo{author}{\bibfnamefont{X.}~\bibnamefont{Chen}},
  \bibinfo{author}{\bibfnamefont{Y.}~\bibnamefont{Chen}},
  \bibinfo{author}{\bibfnamefont{Z.}~\bibnamefont{Cheng}},
  \bibinfo{author}{\bibfnamefont{X.}~\bibnamefont{Cui}},
  \bibinfo{author}{\bibfnamefont{Y.}~\bibnamefont{Fan}},
  \bibinfo{author}{\bibfnamefont{D.}~\bibnamefont{Fang}},
  \bibinfo{author}{\bibfnamefont{Z.}~\bibnamefont{Gao}},
  \bibinfo{author}{\bibfnamefont{L.}~\bibnamefont{Geng}}, \bibnamefont{et~al.}
  (\bibinfo{collaboration}{PandaX Collaboration}),
  \emph{\bibinfo{title}{Measurement of two-neutrino double electron capture
  half-life of $^{124}$xe with pandax-4t}} (\bibinfo{year}{2025}),
  \eprint{2411.14355}, \urlprefix\url{https://arxiv.org/abs/2411.14355}.

\bibitem[{\citenamefont{Aalbers
  et~al.}(2024{\natexlab{c}})\citenamefont{Aalbers, Abe, Adrover, Maouloud,
  Akerib, Musalhi, Alder, Althueser, Amaral, Amarasinghe
  et~al.}}]{XLZD2024_DesignBook}
\bibinfo{author}{\bibfnamefont{J.}~\bibnamefont{Aalbers}},
  \bibinfo{author}{\bibfnamefont{K.}~\bibnamefont{Abe}},
  \bibinfo{author}{\bibfnamefont{M.}~\bibnamefont{Adrover}},
  \bibinfo{author}{\bibfnamefont{S.~A.} \bibnamefont{Maouloud}},
  \bibinfo{author}{\bibfnamefont{D.~S.} \bibnamefont{Akerib}},
  \bibinfo{author}{\bibfnamefont{A.~K.~A.} \bibnamefont{Musalhi}},
  \bibinfo{author}{\bibfnamefont{F.}~\bibnamefont{Alder}},
  \bibinfo{author}{\bibfnamefont{L.}~\bibnamefont{Althueser}},
  \bibinfo{author}{\bibfnamefont{D.~W.~P.} \bibnamefont{Amaral}},
  \bibinfo{author}{\bibfnamefont{C.~S.} \bibnamefont{Amarasinghe}},
  \bibnamefont{et~al.} (\bibinfo{collaboration}{XLZD Collaboration}),
  \emph{\bibinfo{title}{The xlzd design book: Towards the next-generation
  liquid xenon observatory for dark matter and neutrino physics}}
  (\bibinfo{year}{2024}{\natexlab{c}}), \eprint{2410.17137},
  \urlprefix\url{https://arxiv.org/abs/2410.17137}.

\bibitem[{\citenamefont{Migdal}(1941)}]{Migdal_1941}
\bibinfo{author}{\bibfnamefont{A.~B.} \bibnamefont{Migdal}},
  \href{http://dx.doi.org/www.itp.ac.ru/en/persons/migdal-arkady-beinusovich/}{\bibinfo{journal}{J.
  Phys. Acad. Sci. USSR}, \textbf{\bibinfo{volume}{4}},
  \bibinfo{pages}{449}\bibinfo{year}{ (\bibinfo{year}{1941})}}.

\bibitem[{\citenamefont{Ibe et~al.}(2018)\citenamefont{Ibe, Nakano, Shoji, and
  Suzuki}}]{Ibe2017_Migdal}
\bibinfo{author}{\bibfnamefont{M.}~\bibnamefont{Ibe}},
  \bibinfo{author}{\bibfnamefont{W.}~\bibnamefont{Nakano}},
  \bibinfo{author}{\bibfnamefont{Y.}~\bibnamefont{Shoji}}, \bibnamefont{and}
  \bibinfo{author}{\bibfnamefont{K.}~\bibnamefont{Suzuki}},
  \href{http://dx.doi.org/10.1007/JHEP03(2018)194}{\bibinfo{journal}{JHEP},
  \textbf{\bibinfo{volume}{03}}, \bibinfo{pages}{194}\bibinfo{year}{
  (\bibinfo{year}{2018})}}.

\bibitem[{\citenamefont{Akerib et~al.}(2019)\citenamefont{Akerib, Alsum,
  Ara\'ujo, Bai, Balajthy, Beltrame, Bernard, Bernstein, Biesiadzinski, Boulton
  et~al.}}]{LUX2019:Migdal}
\bibinfo{author}{\bibfnamefont{D.~S.} \bibnamefont{Akerib}},
  \bibinfo{author}{\bibfnamefont{S.}~\bibnamefont{Alsum}},
  \bibinfo{author}{\bibfnamefont{H.~M.} \bibnamefont{Ara\'ujo}},
  \bibinfo{author}{\bibfnamefont{X.}~\bibnamefont{Bai}},
  \bibinfo{author}{\bibfnamefont{J.}~\bibnamefont{Balajthy}},
  \bibinfo{author}{\bibfnamefont{P.}~\bibnamefont{Beltrame}},
  \bibinfo{author}{\bibfnamefont{E.~P.} \bibnamefont{Bernard}},
  \bibinfo{author}{\bibfnamefont{A.}~\bibnamefont{Bernstein}},
  \bibinfo{author}{\bibfnamefont{T.~P.} \bibnamefont{Biesiadzinski}},
  \bibinfo{author}{\bibfnamefont{E.~M.} \bibnamefont{Boulton}},
  \bibnamefont{et~al.} (\bibinfo{collaboration}{LUX Collaboration}),
  \href{http://dx.doi.org/10.1103/PhysRevLett.122.131301}{\bibinfo{journal}{Phys.
  Rev. Lett.}, \textbf{\bibinfo{volume}{122}},
  \bibinfo{pages}{131301}\bibinfo{year}{ (\bibinfo{year}{2019})}}.

\bibitem[{\citenamefont{Albakry et~al.}(2023)\citenamefont{Albakry, Alkhatib,
  Alonso-Gonz\'alez, Amaral, Aralis, Aramaki, Arnquist, Ataee~Langroudy,
  Azadbakht, Banik et~al.}}]{CDMS2022_Migdal}
\bibinfo{author}{\bibfnamefont{M.~F.} \bibnamefont{Albakry}},
  \bibinfo{author}{\bibfnamefont{I.}~\bibnamefont{Alkhatib}},
  \bibinfo{author}{\bibfnamefont{D.}~\bibnamefont{Alonso-Gonz\'alez}},
  \bibinfo{author}{\bibfnamefont{D.~W.~P.} \bibnamefont{Amaral}},
  \bibinfo{author}{\bibfnamefont{T.}~\bibnamefont{Aralis}},
  \bibinfo{author}{\bibfnamefont{T.}~\bibnamefont{Aramaki}},
  \bibinfo{author}{\bibfnamefont{I.~J.} \bibnamefont{Arnquist}},
  \bibinfo{author}{\bibfnamefont{I.}~\bibnamefont{Ataee~Langroudy}},
  \bibinfo{author}{\bibfnamefont{E.}~\bibnamefont{Azadbakht}},
  \bibinfo{author}{\bibfnamefont{S.}~\bibnamefont{Banik}}, \bibnamefont{et~al.}
  (\bibinfo{collaboration}{SuperCDMS Collaboration}),
  \href{http://dx.doi.org/10.1103/PhysRevD.107.112013}{\bibinfo{journal}{Phys.
  Rev. D}, \textbf{\bibinfo{volume}{107}},
  \bibinfo{pages}{112013}\bibinfo{year}{ (\bibinfo{year}{2023})}}.

\bibitem[{\citenamefont{Aprile et~al.}(2019{\natexlab{b}})\citenamefont{Aprile,
  Aalbers, Agostini, Alfonsi, Althueser, Amaro, Antochi, Angelino, Arneodo,
  Barge et~al.}}]{XENON1T2019:Migdal}
\bibinfo{author}{\bibfnamefont{E.}~\bibnamefont{Aprile}},
  \bibinfo{author}{\bibfnamefont{J.}~\bibnamefont{Aalbers}},
  \bibinfo{author}{\bibfnamefont{F.}~\bibnamefont{Agostini}},
  \bibinfo{author}{\bibfnamefont{M.}~\bibnamefont{Alfonsi}},
  \bibinfo{author}{\bibfnamefont{L.}~\bibnamefont{Althueser}},
  \bibinfo{author}{\bibfnamefont{F.~D.} \bibnamefont{Amaro}},
  \bibinfo{author}{\bibfnamefont{V.~C.} \bibnamefont{Antochi}},
  \bibinfo{author}{\bibfnamefont{E.}~\bibnamefont{Angelino}},
  \bibinfo{author}{\bibfnamefont{F.}~\bibnamefont{Arneodo}},
  \bibinfo{author}{\bibfnamefont{D.}~\bibnamefont{Barge}}, \bibnamefont{et~al.}
  (\bibinfo{collaboration}{XENON Collaboration}),
  \href{http://dx.doi.org/10.1103/PhysRevLett.123.241803}{\bibinfo{journal}{Phys.
  Rev. Lett.}, \textbf{\bibinfo{volume}{123}},
  \bibinfo{pages}{241803}\bibinfo{year}{
  (\bibinfo{year}{2019}{\natexlab{b}})}}.

\bibitem[{\citenamefont{Agnes et~al.}(2022)\citenamefont{Agnes, Albuquerque,
  Alexander, Alton, Ave, Back, Batignani, Biery, Bocci, Bonivento
  et~al.}}]{DarkSide2022:Migdal}
\bibinfo{author}{\bibfnamefont{P.}~\bibnamefont{Agnes}},
  \bibinfo{author}{\bibfnamefont{I.~F.~M.} \bibnamefont{Albuquerque}},
  \bibinfo{author}{\bibfnamefont{T.}~\bibnamefont{Alexander}},
  \bibinfo{author}{\bibfnamefont{A.~K.} \bibnamefont{Alton}},
  \bibinfo{author}{\bibfnamefont{M.}~\bibnamefont{Ave}},
  \bibinfo{author}{\bibfnamefont{H.~O.} \bibnamefont{Back}},
  \bibinfo{author}{\bibfnamefont{G.}~\bibnamefont{Batignani}},
  \bibinfo{author}{\bibfnamefont{K.}~\bibnamefont{Biery}},
  \bibinfo{author}{\bibfnamefont{V.}~\bibnamefont{Bocci}},
  \bibinfo{author}{\bibfnamefont{W.~M.} \bibnamefont{Bonivento}},
  \bibnamefont{et~al.} (\bibinfo{collaboration}{DarkSide-50 Collaboration}),
  \emph{\bibinfo{title}{Search for dark matter-nucleon interactions via migdal
  effect with darkside-50}} (\bibinfo{year}{2022}),
  \urlprefix\url{https://arxiv.org/abs/2207.11967}.

\bibitem[{\citenamefont{Xu et~al.}(2024)\citenamefont{Xu, Adams, Lenardo,
  Pershing, Mannino, Bernard, Kingston, Mizrachi, Lin, Essig
  et~al.}}]{Xu2024_Migdal}
\bibinfo{author}{\bibfnamefont{J.}~\bibnamefont{Xu}},
  \bibinfo{author}{\bibfnamefont{D.}~\bibnamefont{Adams}},
  \bibinfo{author}{\bibfnamefont{B.}~\bibnamefont{Lenardo}},
  \bibinfo{author}{\bibfnamefont{T.}~\bibnamefont{Pershing}},
  \bibinfo{author}{\bibfnamefont{R.}~\bibnamefont{Mannino}},
  \bibinfo{author}{\bibfnamefont{E.}~\bibnamefont{Bernard}},
  \bibinfo{author}{\bibfnamefont{J.}~\bibnamefont{Kingston}},
  \bibinfo{author}{\bibfnamefont{E.}~\bibnamefont{Mizrachi}},
  \bibinfo{author}{\bibfnamefont{J.}~\bibnamefont{Lin}},
  \bibinfo{author}{\bibfnamefont{R.}~\bibnamefont{Essig}},
  \bibnamefont{et~al.},
  \href{http://dx.doi.org/https://doi.org/10.1103/PhysRevD.109.L051101}{\bibinfo{journal}{Physical
  Review D}, \textbf{\bibinfo{volume}{109}},
  \bibinfo{pages}{L051101}\bibinfo{year}{ (\bibinfo{year}{2024})}}.

\bibitem[{\citenamefont{Bang}(2024)}]{LZ_Bang_phd}
\bibinfo{author}{\bibfnamefont{J.}~\bibnamefont{Bang}}, Ph.D. thesis,
  \bibinfo{school}{Brown University} (\bibinfo{year}{2024}),
  \urlprefix\url{https://repository.library.brown.edu/studio/item/bdr:37s8v2v2/}.

\bibitem[{\citenamefont{Vaitkus}(2025)}]{LZ_Vaitkus_phd}
\bibinfo{author}{\bibfnamefont{A.}~\bibnamefont{Vaitkus}}, Ph.D. thesis,
  \bibinfo{school}{Brown University} (\bibinfo{year}{2025}),
  \urlprefix\url{https://repository.library.brown.edu/studio/item/bdr:7jaat7fh/}.

\bibitem[{\citenamefont{Aalbers et~al.}(2023)\citenamefont{Aalbers, Akerib,
  Musalhi, Alder, Alsum, Amarasinghe, Ames, Anderson, Angelides, Ara\'ujo
  et~al.}}]{LZ2023_Background}
\bibinfo{author}{\bibfnamefont{J.}~\bibnamefont{Aalbers}},
  \bibinfo{author}{\bibfnamefont{D.~S.} \bibnamefont{Akerib}},
  \bibinfo{author}{\bibfnamefont{A.~K.~A.} \bibnamefont{Musalhi}},
  \bibinfo{author}{\bibfnamefont{F.}~\bibnamefont{Alder}},
  \bibinfo{author}{\bibfnamefont{S.~K.} \bibnamefont{Alsum}},
  \bibinfo{author}{\bibfnamefont{C.~S.} \bibnamefont{Amarasinghe}},
  \bibinfo{author}{\bibfnamefont{A.}~\bibnamefont{Ames}},
  \bibinfo{author}{\bibfnamefont{T.~J.} \bibnamefont{Anderson}},
  \bibinfo{author}{\bibfnamefont{N.}~\bibnamefont{Angelides}},
  \bibinfo{author}{\bibfnamefont{H.~M.} \bibnamefont{Ara\'ujo}},
  \bibnamefont{et~al.} (\bibinfo{collaboration}{The LUX-ZEPLIN Collaboration}),
   \href{http://dx.doi.org/10.1103/PhysRevD.108.012010}{\bibinfo{journal}{Phys.
  Rev. D}, \textbf{\bibinfo{volume}{108}},
  \bibinfo{pages}{012010}\bibinfo{year}{ (\bibinfo{year}{2023})}}.

\bibitem[{\citenamefont{Fitzpatrick et~al.}(2013)\citenamefont{Fitzpatrick,
  Haxton, Katz, Lubbers, and Xu}}]{Fitzpatrick:2012ix}
\bibinfo{author}{\bibfnamefont{A.~L.} \bibnamefont{Fitzpatrick}},
  \bibinfo{author}{\bibfnamefont{W.}~\bibnamefont{Haxton}},
  \bibinfo{author}{\bibfnamefont{E.}~\bibnamefont{Katz}},
  \bibinfo{author}{\bibfnamefont{N.}~\bibnamefont{Lubbers}}, \bibnamefont{and}
  \bibinfo{author}{\bibfnamefont{Y.}~\bibnamefont{Xu}},
  \href{http://dx.doi.org/10.1088/1475-7516/2013/02/004}{\bibinfo{journal}{JCAP},
  \textbf{\bibinfo{volume}{02}}, \bibinfo{pages}{004}\bibinfo{year}{
  (\bibinfo{year}{2013})}}.

\bibitem[{\citenamefont{Thomas and Imel}(1987)}]{Thomas1987_Recombination}
\bibinfo{author}{\bibfnamefont{J.}~\bibnamefont{Thomas}} \bibnamefont{and}
  \bibinfo{author}{\bibfnamefont{D.}~\bibnamefont{Imel}},
  \href{http://dx.doi.org/https://doi.org/10.1103/PhysRevA.36.614}{\bibinfo{journal}{Physical
  Review A}, \textbf{\bibinfo{volume}{36}}, \bibinfo{pages}{614}\bibinfo{year}{
  (\bibinfo{year}{1987})}}.

\bibitem[{\citenamefont{Mozumder}(1995)}]{Mozumder1995_Recombination}
\bibinfo{author}{\bibfnamefont{A.}~\bibnamefont{Mozumder}},
  \href{http://dx.doi.org/https://doi.org/10.1016/0009-2614(95)01024-4}{\bibinfo{journal}{Chemical
  Physics Letters}, \textbf{\bibinfo{volume}{245}},
  \bibinfo{pages}{359}\bibinfo{year}{ (\bibinfo{year}{1995})}}, ISSN
  \bibinfo{issn}{0009-2614}.

\bibitem[{\citenamefont{Takahashi et~al.}(1975)\citenamefont{Takahashi, Konno,
  Hamada, Miyajima, Kubota, Nakamoto, Hitachi, Shibamura, and
  Doke}}]{Takahashi1975_Xe}
\bibinfo{author}{\bibfnamefont{T.}~\bibnamefont{Takahashi}},
  \bibinfo{author}{\bibfnamefont{S.}~\bibnamefont{Konno}},
  \bibinfo{author}{\bibfnamefont{T.}~\bibnamefont{Hamada}},
  \bibinfo{author}{\bibfnamefont{M.}~\bibnamefont{Miyajima}},
  \bibinfo{author}{\bibfnamefont{S.}~\bibnamefont{Kubota}},
  \bibinfo{author}{\bibfnamefont{A.}~\bibnamefont{Nakamoto}},
  \bibinfo{author}{\bibfnamefont{A.}~\bibnamefont{Hitachi}},
  \bibinfo{author}{\bibfnamefont{E.}~\bibnamefont{Shibamura}},
  \bibnamefont{and} \bibinfo{author}{\bibfnamefont{T.}~\bibnamefont{Doke}},
  \href{http://dx.doi.org/10.1103/PhysRevA.12.1771}{\bibinfo{journal}{Phys.
  Rev. A}, \textbf{\bibinfo{volume}{12}}, \bibinfo{pages}{1771}\bibinfo{year}{
  (\bibinfo{year}{1975})}}.

\bibitem[{\citenamefont{Doke et~al.}(2002)\citenamefont{Doke, Hitachi, Kikuchi,
  Masuda, Okada, and Shibamura}}]{Doke2002_ArXe}
\bibinfo{author}{\bibfnamefont{T.}~\bibnamefont{Doke}},
  \bibinfo{author}{\bibfnamefont{A.}~\bibnamefont{Hitachi}},
  \bibinfo{author}{\bibfnamefont{J.}~\bibnamefont{Kikuchi}},
  \bibinfo{author}{\bibfnamefont{K.}~\bibnamefont{Masuda}},
  \bibinfo{author}{\bibfnamefont{H.}~\bibnamefont{Okada}}, \bibnamefont{and}
  \bibinfo{author}{\bibfnamefont{E.}~\bibnamefont{Shibamura}},
  \href{http://stacks.iop.org/1347-4065/41/i=3R/a=1538}{\bibinfo{journal}{Japanese
  Journal of Applied Physics}, \textbf{\bibinfo{volume}{41}},
  \bibinfo{pages}{1538}\bibinfo{year}{ (\bibinfo{year}{2002})}}.

\bibitem[{\citenamefont{Niţescu et~al.}(2024)\citenamefont{Niţescu, Ghinescu,
  Sevestrean, Horoi, Šimkovic, and Stoica}}]{Nitescu2024_124Xe}
\bibinfo{author}{\bibfnamefont{O.}~\bibnamefont{Niţescu}},
  \bibinfo{author}{\bibfnamefont{S.}~\bibnamefont{Ghinescu}},
  \bibinfo{author}{\bibfnamefont{V.~A.} \bibnamefont{Sevestrean}},
  \bibinfo{author}{\bibfnamefont{M.}~\bibnamefont{Horoi}},
  \bibinfo{author}{\bibfnamefont{F.}~\bibnamefont{Šimkovic}},
  \bibnamefont{and} \bibinfo{author}{\bibfnamefont{S.}~\bibnamefont{Stoica}},
  \href{http://dx.doi.org/10.1088/1361-6471/ad8767}{\bibinfo{journal}{Journal
  of Physics G: Nuclear and Particle Physics}, \textbf{\bibinfo{volume}{51}},
  \bibinfo{pages}{125103}\bibinfo{year}{ (\bibinfo{year}{2024})}}.

\bibitem[{\citenamefont{Baudis et~al.}(2021)\citenamefont{Baudis,
  Sanchez-Lucas, and Thieme}}]{Baudis2021_WValue}
\bibinfo{author}{\bibfnamefont{L.}~\bibnamefont{Baudis}},
  \bibinfo{author}{\bibfnamefont{P.}~\bibnamefont{Sanchez-Lucas}},
  \bibnamefont{and} \bibinfo{author}{\bibfnamefont{K.}~\bibnamefont{Thieme}},
  \href{http://dx.doi.org/10.1140/epjc/s10052-021-09834-x}{\bibinfo{journal}{The
  European Physical Journal C}, \textbf{\bibinfo{volume}{81}},
  \bibinfo{pages}{1060}\bibinfo{year}{ (\bibinfo{year}{2021})}}, ISSN
  \bibinfo{issn}{1434-6052}.

\bibitem[{\citenamefont{Anton et~al.}(2020)\citenamefont{Anton, Badhrees,
  Barbeau, Beck, Belov, Bhatta, Breidenbach, Brunner, Cao, Cen
  et~al.}}]{EXO2020_WValue}
\bibinfo{author}{\bibfnamefont{G.}~\bibnamefont{Anton}},
  \bibinfo{author}{\bibfnamefont{I.}~\bibnamefont{Badhrees}},
  \bibinfo{author}{\bibfnamefont{P.~S.} \bibnamefont{Barbeau}},
  \bibinfo{author}{\bibfnamefont{D.}~\bibnamefont{Beck}},
  \bibinfo{author}{\bibfnamefont{V.}~\bibnamefont{Belov}},
  \bibinfo{author}{\bibfnamefont{T.}~\bibnamefont{Bhatta}},
  \bibinfo{author}{\bibfnamefont{M.}~\bibnamefont{Breidenbach}},
  \bibinfo{author}{\bibfnamefont{T.}~\bibnamefont{Brunner}},
  \bibinfo{author}{\bibfnamefont{G.~F.} \bibnamefont{Cao}},
  \bibinfo{author}{\bibfnamefont{W.~R.} \bibnamefont{Cen}},
  \bibnamefont{et~al.} (\bibinfo{collaboration}{EXO-200 Collaboration}),
  \href{http://dx.doi.org/10.1103/PhysRevC.101.065501}{\bibinfo{journal}{Phys.
  Rev. C}, \textbf{\bibinfo{volume}{101}},
  \bibinfo{pages}{065501}\bibinfo{year}{ (\bibinfo{year}{2020})}}.

\bibitem[{\citenamefont{Perkins et~al.}(1991)\citenamefont{Perkins, Cullen,
  Chen, Hubbell, Rathkopf, and Scofield}}]{perkins1991eadl}
\bibinfo{author}{\bibfnamefont{S.}~\bibnamefont{Perkins}},
  \bibinfo{author}{\bibfnamefont{D.}~\bibnamefont{Cullen}},
  \bibinfo{author}{\bibfnamefont{M.}~\bibnamefont{Chen}},
  \bibinfo{author}{\bibfnamefont{J.}~\bibnamefont{Hubbell}},
  \bibinfo{author}{\bibfnamefont{J.}~\bibnamefont{Rathkopf}}, \bibnamefont{and}
  \bibinfo{author}{\bibfnamefont{J.}~\bibnamefont{Scofield}},
  \href{https://inis.iaea.org/records/cyywt-mds55}{\bibinfo{journal}{Lawrence
  Livermore National Laboratory, UCRL},
  \textbf{\bibinfo{volume}{50400}}\bibinfo{year}{ (\bibinfo{year}{1991})}}.

\end{thebibliography}

\end{document}